\documentclass[12pt,dvips]{article}

\usepackage{rotating,amsmath}
\usepackage{epsfig}
\usepackage{color}
\usepackage{cite}

\usepackage{hhline}
\usepackage{multirow}

\numberwithin{equation}{section}

\newcommand{\lsim}{\raisebox{-0.13cm}{~\shortstack{$<$ \\[-0.07cm] $\sim$}}~}
\newcommand{\gsim}{\raisebox{-0.13cm}{~\shortstack{$>$ \\[-0.07cm] $\sim$}}~}

\begin{document}

\def\thefootnote{\fnsymbol{footnote}}

\begin{flushright}
{\tt MAN/HEP/2012/01, 
FTUV-12-2401}\\
{\tt arXiv:1201.4891} \\
January 2012
\end{flushright}

\medskip

\begin{center}
{\bf {\Large Radiative Corrections to Scalar Masses and Mixing \\[3mm]
    in a Scale Invariant Two Higgs Doublet Model}}
\end{center}

\medskip

\begin{center}{\large
Jae~Sik~Lee$^{a,b,c}$ and Apostolos~Pilaftsis$^{c,d}$}
\end{center}

\begin{center}
{\it $^a$Department of Physics, National Tsing Hua University,
Hsinchu, Taiwan 300}\\[0.2cm]
{\it $^b$ Department of Physics, Chonnam National University, \\
300 Yongbong-dong, Buk-gu, Gwangju,
500-757, Republic of Korea }\\[2mm]
{\it $^c$Consortium for Fundamental Physics, School of Physics
and Astronomy,\\ University of
Manchester, Manchester M13 9PL, United Kingdom}\\[2mm]
{\it $^d$Department of Theoretical Physics and
    IFIC, University of Valencia--CSIC,\\ E-46100, Valencia, Spain}
\end{center}

\bigskip\bigskip

\centerline{\bf ABSTRACT}

\noindent  
We study the Higgs-boson  mass spectrum of a classical scale-invariant
realization of  the two-Higgs-doublet model  (SI-2HDM).  The classical
scale  symmetry of  the theory  is explicitly  broken by  quantum loop
effects due to gauge  interactions, Higgs self-couplings and top-quark
Yukawa couplings.  We determine the allowed parameter space compatible
with  perturbative unitarity and  electroweak precision  data.  Taking
into  account  the  LEP and  the  recent  LHC  exclusion limits  on  a
Standard-Model-like Higgs boson $H_{\rm  SM}$, we obtain rather strict
constraints  on  the  mass  spectrum  of the  heavy  Higgs  sector  of
the~SI-2HDM.   In~particular, if  $M_{H_{\rm SM}}  \sim  125$~GeV, the
SI-2HDM  strongly favours  scenarios, in  which  at least  one of  the
non-standard neutral Higgs  bosons has a mass close to  400 GeV and is
generically degenerate with the  charged Higgs boson, whilst the third
neutral Higgs scalar is lighter than $\sim 500$~GeV.


%
%

\newpage

\section{Introduction}
\label{sec:intro}

Classical scale  symmetries provide a minimal  and calculable approach
to potentially  solving the infamous gauge hierarchy  problem.  In the
Standard Model (SM), the absence  of the mass parameter~$m^2$ from the
Higgs  potential renders  the  classical action  of  the theory  scale
invariant~(SI).  However,  as originally  discussed by Coleman  and E.
Weinberg \cite{Coleman:1973jx} and later  by Gildener and S.  Weinberg
\cite{Gildener:1976ih}, quantum loops generate logarithmic terms which
anomalously break the  scale invariance of the theory,  giving rise to
electroweak symmetry  breaking.  Given the  LEP2 mass limit on  the SM
Higgs             boson             $M_{H_{\mathrm{SM}}}             >
114.4~\mathrm{GeV}$~\cite{Barate:2003sz} and the experimental value of
the top-quark mass $m_t \approx 173$~GeV, a perturbative SI version of
the SM  is not both  theoretically and phenomenologically  viable.  In
particular,  the large  top-quark  Yukawa coupling  gives  rise to  an
effective potential which is no longer bounded from below, at least at
the  perturbative  level.  This  difficulty  may  be circumvented,  if
additional massive bosonic fields 
such      as       real       and      complex       singlet
scalars are  present in SI extensions of 
the SM~\cite{Hempfling:1996ht,Chang:2007ki,Iso:2009ss,Foot:2007as,
MNmar2007,AlexanderNunneley:2010nw,Holthausen:2009uc}.

In  this paper  we study  a minimal  Scale-Invariant two-Higgs-Doublet
Model   (SI-2HDM)   extension  of   the   SM.    To  naturally   avoid
flavour-changing neutral currents (FCNCs),  we assume that the SI-2HDM
potential    is    invariant   under    a    ${\rm   Z}_2$    discrete
symmetry~\cite{Glashow:1976nt},  under which  the  two Higgs  doublets
$\Phi_{1,2}$ transform  as $\Phi_{1  (2)} \to +(-)  \Phi_{1\,(2)}$. At
the  tree  level, the  spontaneous  breaking  of  the classical  scale
symmetry due to the presence  of a non-vanishing flat direction in the
Higgs  potential gives  rise  to a  massless CP-even  pseudo-Goldstone
boson~$h$.   We calculate  the  radiative corrections  to the  CP-even
Higgs-boson mass matrix that result  from quantum loops of $W^\pm$ and
$Z$  bosons, Higgs self-interactions  and top-quark  Yukawa couplings.
To determine the  allowed parameter space of the  SI-2HDM, we consider
the theoretical  constraints of convexity  and perturbative unitarity,
as  well as  phenomenological constraints  from  electroweak precision
data and direct Higgs-boson searches.

Taking all  the aforementioned  constraints into account,  the allowed
range of  masses for the  charged Higgs bosons~$H^\pm$ and  the CP-odd
scalar $A$ gets significantly restricted.   We find that for a 125-GeV
SM-like  Higgs  boson  $H_1$,  at  least  two  Higgs  states,  charged
($H^\pm$) or  neutral ($H_2,A$),  are generically degenerate  and have
masses  close to  400~GeV, whereas  the third  Higgs state  has  to be
lighter  than  500~GeV.  In  particular,  there  are three  favourable
scenarios with the above  characteristics.  In the first scenario, the
CP-even  Higgs  boson  $H_2$  and  the CP-odd  scalar~$A$  are  almost
degenerate with $M_{H_2} \sim M_A \sim 400$~GeV, and the charged Higgs
boson~$H^\pm$  weighs between  $295$ GeV  and $420$~GeV,  after taking
into account  the $b  \to s \gamma$  constraint.  The  second favourable
scenario contains  a CP-odd  state $A$ lighter  than 100 GeV,  and the
Higgs  states  $H^\pm$  and  $H_2$  have  approximately  equal  masses
$M_{H^\pm}\sim  M_{H_2}  \sim 400$~GeV.   Finally,  there  is a  third
possibility,  where  the heavier  CP-even  Higgs  boson  $H_2$ can  be
lighter than 180  GeV, while the charged Higgs  bosons $H^\pm$ and the
CP-odd  scalar  $A$  are  restricted  to be  almost  degenerate,  with
$M_{H^\pm}\sim M_{A} \sim 400$~GeV.

The layout of the paper  is as follows. After this brief introduction,
in Section~\ref{sec:SI-2HDM} we discuss  in detail the Higgs sector of
the SI-2HDM.  Specifically, we  first determine the flat directions of
the  tree-level SI-2HDM potential  and its  scalar mass  spectrum.  We
then  calculate the one-loop  effective potential  of the  SI-2HDM and
evaluate the radiatively corrected  masses of the CP-even Higgs bosons
and  their  mixing.   At the  end  of  this  section, we  discuss  the
importance  of  the  choice of  the  RG  scale  in our  analysis.   In
Section~\ref{sec:analysis}  we impose  the  theoretical constraint  of
perturbative unitarity and require compatibility of the theory against
electroweak precision  data and  direct Higgs-boson searches.   In the
light of these restrictions,  we determine the allowed parameter space
for   the    heavy   Higgs   sector   of    the   SI-2HDM.    Finally,
Section~\ref{sec:conclusions} summarizes our conclusions and discusses
possible future directions.

\section{Scale Invariant Two Higgs Doublet Model}\label{sec:SI-2HDM}

The 2HDM exhibits an exact classical scaling symmetry, if there are no
explicit  mass parameters  in the  scalar potential.  To be specific,
under global scale transformations:
\begin{equation}
  \label{eq:ST}
\varphi (x)\quad \to \quad \varphi' (x')\ =\  e^{d_\varphi \sigma}\, 
\varphi ( e^\sigma x)\; ,
\end{equation}
where  $\sigma$ is  a  constant,  the action  of  the 2HDM  Lagrangian
$S[\varphi (x)]$ remains  invariant, i.e.~$S[\varphi (x)] = S[\varphi'
  (x')]$,  where $\varphi$  represents a  generic  bosonic (fermionic)
field of the  2HDM and $d_\varphi = 1$~(3/2)  is its classical scaling
dimension. Beyond  the tree level,  the classical scale  invariance of
the  theory  is  broken   by  scalar  operators  of  dimension  $n>4$,
e.g.~$\varphi^4   \ln   (\varphi^2/\langle   \varphi\rangle^2)$   with
$d_\varphi = 1$, in a  SI (or~no-scale) regularization scheme, such as
the      scheme      of      dimensional      regularization      (see
also~\cite{AlexanderNunneley:2010nw}, and references therein). This is
the  scheme that  we consider  here  for performing  our quantum  loop
calculations. Nevertheless,  had we chosen  a scheme with  explicit UV
cut-off  dependence,  we  would  have  obtained the  same  results  by
demanding that the  renormalized Coleman--Weinberg effective potential
$V_{\rm  eff}$  satisfies  the  conditions: $d^n  V_{\rm  eff}(\varphi
)/d\varphi^n = 0$ at $\varphi = 0$, for $n =0,1,2,3$.

We note that our approach to formulating a classical SI theory differs
from the one studied  in~\cite{Misha,GGS}, where the scale symmetry is
imposed at  the quantum level.   As~argued in~\cite{Tkachov}, however,
quantum SI  theories face difficulties with  renormalizability at high
orders  and they  can therefore  be regarded  only as  effective field
theories.

In this  section, after introducing the  tree-level SI-2HDM potential,
we  determine  its  flat  directions  and the  resulting  scalar  mass
spectrum. Then,  we calculate  the one-loop effective  potential, from
which  we   derive  the  radiatively   corrected  Higgs-boson  masses.
Finally, we  comment on the  choice of the  renormalization-group (RG)
scale.

\subsection{Flat Directions of the Tree-Level Potential}\label{sec:tree}

At the tree-level, the most general SI-2HDM potential reads:
\begin{eqnarray}
  \label{eq:V2HDM}
\mathrm{V}^0 \!& = &\! \lambda_1 (\Phi_1^{\dagger} \Phi_1)^2 +
\lambda_2 (\Phi_2^{\dagger} \Phi_2)^2\: +\: \lambda_3
(\Phi_1^{\dagger} 
\Phi_1)(\Phi_2^{\dagger} \Phi_2) + \lambda_4 (\Phi_1^{\dagger}
\Phi_2)(\Phi_2^{\dagger} \Phi_1)\nonumber\\  
\!&&\! +\: \frac{\lambda_5}{2}
(\Phi_1^{\dagger} \Phi_2)^2 +  \frac{\lambda_5^{*}}{2}
(\Phi_2^{\dagger} \Phi_1)^2 + 
\lambda_6 (\Phi_1^{\dagger} \Phi_1) (\Phi_1^{\dagger} \Phi_2) + \lambda_6^{*}
(\Phi_1^{\dagger} \Phi_1)(\Phi_2^{\dagger} \Phi_1)\nonumber\\
\!&&\! +\: \lambda_7 (\Phi_2^{\dagger} \Phi_2) (\Phi_1^{\dagger} \Phi_2)\: +\:
\lambda_7^{*} (\Phi_2^{\dagger} \Phi_2) (\Phi_2^{\dagger} \Phi_1)\; .
\end{eqnarray}
In order to  naturally avoid too large FCNC  interactions of the Higgs
bosons   to    quarks,   we   impose   the    ${\rm   Z}_2$   discrete
symmetry~\cite{Glashow:1976nt}: $\Phi_{1\,(2)} \to +(-) \Phi_{1\,(2)}$
(for a  recent review see~\cite{Branco}). In such  a minimal scenario,
the  quartic couplings  $\lambda_6$  and $\lambda_7$  vanish, and  the
CP-odd  phase of  $\lambda_5$ can  be rotated  away, i.e.~there  is no
explicit CP violation at the tree level.

Assuming that  only the neutral  components of the two  Higgs doublets
$\Phi_{1,2}$ develop  non-vanishing vacuum expectation  values (VEVs),
we may parameterize $\Phi_{1,2}$ as follows:
\begin{equation}
\Phi_1=\left(\begin{array}{c}
\phi_1^+ \\ \frac{1}{\sqrt{2}}\,(v_1+\phi_1+ia_1)
\end{array}\right)\;, \qquad
\Phi_2=
\left(\begin{array}{c}
\phi_2^+ \\ \frac{1}{\sqrt{2}}\,(v_2+\phi_2+ia_2)
\end{array}\right)\, .
\end{equation}
We   denote  $v_1\equiv   v  \cos\beta=vc_\beta$   and   $v_2\equiv  v
\sin\beta=vs_\beta$, where  $v \simeq  246$~GeV is the  VEV of  the SM
Higgs doublet. Extremizing the tree-level scalar potential~$V^0$ leads
to the following tadpole conditions:
\begin{eqnarray}
T_{\phi_1}\ \equiv\ \left\langle\frac{\partial V^0}{\partial
  \phi_1}\right\rangle &=& 
v_1 \left(\lambda_1 v_1^2+\frac{1}{2}\lambda_{345} v_2^2 \right)
\  =\ 0\;,
\nonumber \\[2mm]
T_{\phi_2}\ \equiv\ \left\langle\frac{\partial V^0}{\partial
  \phi_2}\right\rangle &=& 
v_2 \left(\lambda_2 v_2^2+\frac{1}{2}\lambda_{345} v_1^2 \right)
\ =\ 0\;,
\end{eqnarray}
with    $\lambda_{345}\equiv    \lambda_3+\lambda_4+\lambda_5$.    The
vanishing  of  the  tadpole  parameters $T_{\phi_{1,2}}$  is  ensured,
provided
\begin{equation}
  \label{eq:tad0}
\frac{\lambda_1}{\lambda_2}\ =\ \tan^4\beta \;, \qquad
2 \sqrt{\lambda_1\lambda_2}\ =\ \pm \lambda_{345}\; .
\end{equation}
As we will see below, requiring a convex, bounded-from-below potential
and a non-negative scalar mass  spectrum fixes the $\pm$ sign in front
of $\lambda_{345}$, which turns out to be minus.

In detail,  the tree-level  mass spectrum of  the charged  and neutral
Higgs bosons may be calculated as
\begin{eqnarray}
V^0_{\rm mass} &=&
\Big(G^+\,,  \ H^+\Big)\left(\begin{array}{cc}
0 & 0 \\ 0 & M_{H^\pm}^2 \end{array}\right)
\left(\begin{array}{c}
G^- \\ H^- \end{array}\right) \ + \ \frac{1}{2}\:
\Big(G^0\,,  \ A\Big)\left(\begin{array}{cc}
0 & 0 \\ 0 & M_A^2 \end{array}\right)
\left(\begin{array}{c}
G^0 \\ A \end{array}\right) 
\nonumber \\[3mm]
&& + \ \frac{v^2}{2}\:
\Big(\phi_1\,, \ \phi_2 \Big)\,
\left(\begin{array}{cc}
2\lambda_1\, c_\beta^2 & \lambda_{345}\,c_\beta s_\beta \\[1mm]
\lambda_{345}\,c_\beta s_\beta & 2\lambda_2\, s_\beta^2
\end{array}\right)\,
\left(\begin{array}{c}
\phi_1 \\ \phi_2  \end{array}\right)\; ,
\end{eqnarray}
where
\begin{equation}
\left(\begin{array}{c} \phi_1^- \\ \phi_2^- \end{array}\right) =
\left(\begin{array}{rr}
c_\beta & -s_\beta \\ s_\beta & c_\beta \end{array}\right)\,
\left(\begin{array}{c} G^- \\ H^- \end{array}\right) \;, \qquad
\left(\begin{array}{c} a_1 \\ a_2 \end{array}\right) =
\left(\begin{array}{rr}
c_\beta & -s_\beta \\ s_\beta & c_\beta \end{array}\right)\,
\left(\begin{array}{c} G^0 \\ A \end{array}\right) \;
\end{equation}
and  
\begin{equation}
  \label{eq:MHA0}
M_{H^\pm}^2\ =\ -\frac{1}{2}\,(\lambda_4+\lambda_5)\,v^2 \;, \qquad
M_A^2\ =\ -\lambda_5\,v^2 
\end{equation}
are the squared masses of the charged and CP-odd Higgs bosons, $H^\pm$
and $A$, respectively. In addition, we observe that the determinant of
the $2\times 2$ CP-even  Higgs-boson mass matrix vanishes identically,
as a consequence of the second tadpole condition in~(\ref{eq:tad0}).

The  vanishing of  the  determinant of  the  CP-even Higgs-boson  mass
matrix  signifies   the  existence  of   a  massless  pseudo-Goldstone
boson~$h$,  arising  from  the  spontaneous breaking  of  the  scaling
symmetry along a  minimal flat direction of the  SI-2HDM potential. In
order  to determine the  flat direction,  we perform an orthogonal
transformation on the CP-even scalar fields:
\begin{equation}
\left(\begin{array}{c} \phi_1 \\ \phi_2 \end{array}\right)\ =\
\left(\begin{array}{rr}
c_\alpha & -s_\alpha \\ s_\alpha & c_\alpha \end{array}\right)\,
\left(\begin{array}{c} H \\ h \end{array}\right) \; ,
\end{equation}
so as to render the CP-even scalar mass matrix diagonal, i.e.
\begin{equation}
\left(\begin{array}{rr}
c_\alpha & s_\alpha \\ -s_\alpha & c_\alpha \end{array}\right)\,
\left(\begin{array}{cc}
2\lambda_1\, c_\beta^2 & \lambda_{345}\,c_\beta s_\beta \\[1mm]
\lambda_{345}\,c_\beta s_\beta & 2\lambda_2\, s_\beta^2
\end{array}\right)\,
\left(\begin{array}{rr}
c_\alpha & -s_\alpha \\ s_\alpha & c_\alpha \end{array}\right)\ = \
\left(\begin{array}{cc}
M_H^2/v^2 & 0 \\ 0 & 0 \end{array}\right)\; .
\end{equation}
In this way, we obtain 
\begin{eqnarray}
  \label{eq:MH0}
M_H^2\  =\  -\lambda_{345}\,v^2\ =\ 2\sqrt{\lambda_1\lambda_2}\,v^2\;, \qquad 
\sin^2(\alpha-\beta)\ =\ 1\; .
\end{eqnarray}
Observe that  positivity of $M^2_H$ requires to  have $\lambda_{345} <
0$.   Moreover, the coupling  of the  massive state  ($H$) to  the two
vector bosons vanishes,  while the coupling of the  massless state $h$
is the same as the SM one $H_{\rm SM}$:
\begin{equation}
\frac{g_{HWW}^2}{g_{H_{\rm SM}WW}^2}\ =\ \cos^2(\alpha-\beta)\ =\ 0\;,\qquad
\frac{g_{hWW}^2}{g_{H_{\rm SM}WW}^2}\ =\ \sin^2(\alpha-\beta)\ =\ 1\; .
\end{equation}
The  flat direction  $\phi_{\rm  Flat}$ associated  with the  massless
CP-even scalar $h$  may be expressed in different  equivalent forms as
follows:
\begin{equation}
\phi_{\rm Flat}\ =\ v +h \ =\ v-s_\alpha \phi_1 + c_\alpha \phi_2\
=\ c_\beta (v_1+\phi_1) + s_\beta (v_2+\phi_2)\; ,
\end{equation}
where  we take  $\langle\phi_{\rm  Flat} \rangle  =  v$ and  $s_\alpha
=-c_\beta $ and $c_\alpha =s_\beta$.

In  summary, gathering the  results derived  above in~(\ref{eq:tad0}),
(\ref{eq:MHA0}) and (\ref{eq:MH0}), we have the following constraining
set of input parameters:
\begin{eqnarray}
  \label{eq:set2}
t_\beta^2 \!&=&\! \sqrt{\frac{\lambda_1}{\lambda_2}} \ , \qquad
M_H^2\ =\ -(\lambda_3+\lambda_4+\lambda_5)\, v^2 \ =\ 
2\sqrt{\lambda_1\lambda_2}\, v^2\; , \nonumber\\
M_{H^\pm}^2 \!&=&\! -\frac{1}{2}(\lambda_4+\lambda_5)\, v^2 \; ,\qquad
M_A^2\ =\ -\lambda_5\, v^2\; .
\end{eqnarray}
Note  that all  the three  tree-level Higgs  masses can  be determined
entirely  by   the  three  couplings   $\lambda_3$,  $\lambda_4$,  and
$\lambda_5$ and  the SM  VEV $v$, independently  of $t_\beta$.  We may
also invert  the relations given in~(\ref{eq:set2})  and determine the
five  quartic  couplings   $\lambda_{1,2,3,4,5}$,  in  terms  of  $v$,
$t_\beta$, and the three Higgs masses:
\begin{eqnarray}
  \label{eq:set1}
\lambda_1 \!& = &\! \frac{M_H^2}{2v^2}\,t_\beta^2 \ , \qquad
\lambda_2\  =\ \frac{M_H^2}{2v^2\, t_\beta^2} \ , \nonumber\\
\lambda_3  \!& = &\! \frac{2M_{H^\pm}^2-M_H^2}{v^2} \ , \qquad
\lambda_4\ =\ \frac{M_A^2-2M_{H^\pm}^2}{v^2} \ , \qquad
\lambda_5\ =\ -\frac{M_A^2}{v^2}\ .
\end{eqnarray}

Finally, it is  interesting to comment on the  convexity conditions of
the               ${\rm              Z}_2$-invariant              2HDM
potential~\cite{Deshpande:1977rw,ElKaffas:2006nt}.  These are given by
\begin{equation}
\lambda_1\ >\ 0\;,\qquad  \lambda_2\ >\ 0\;, \qquad
2\sqrt{\lambda_1\lambda_2}\: +\: \lambda_3\: +\:
{\rm min} \left[0\,,\lambda_4+\lambda_5\,,\lambda_4-\lambda_5\right]\ >\ 0\; .
\end{equation}
While the first  two conditions are easily satisfied,  we observe that
the  third expression  of  the couplings  vanishes identically,  since
${\rm  min} \left[0\,,\lambda_4+\lambda_5\,,\lambda_4-\lambda_5\right]
=\lambda_4+\lambda_5$, and  $\lambda_3 + \lambda_4  + \lambda_5 =  - 2
\sqrt{\lambda_1  \lambda_2}$~[cf.~(\ref{eq:tad0})].  The  vanishing of
the third expression signals the  existence of a flat direction in the
SI-2HDM potential,  which gets lifted  by radiative corrections  as we
discuss below.

\subsection{One-Loop Effective Potential}\label{sec:rad}

As mentioned above, it is important to consider the quantum effects on
the tree-level potential. More explicitly, the one-loop effective 
potential~\cite{Coleman:1973jx} may be calculated as 
\begin{eqnarray}
V_{\rm eff}^{\rm 1-loop} \!&=&\! \frac{1}{64\pi^2}\left[
 M_H^4\left(-\frac{3}{2}+\ln\frac{M_H^2}{Q^2}\right)
+M_A^4\left(-\frac{3}{2}+\ln\frac{M_A^2}{Q^2}\right)
+2M_{H^\pm}^4\left(-\frac{3}{2}+\ln\frac{M_{H^\pm}^2}{Q^2}\right) \right.
\nonumber \\[3mm] 
&& \hspace{-0.5cm} \left.
+\: 6M_W^4\left(-\frac{5}{6}+\ln\frac{M_W^2}{Q^2}\right)
+3M_Z^4\left(-\frac{5}{6}+\ln\frac{M_Z^2}{Q^2}\right)
-12m_t^4\left(-1+\ln\frac{m_t^2}{Q^2}\right) \right] ,
\label{eq:1lpot}
\end{eqnarray}
where $Q$  is the RG  scale and the background  field-dependent masses
are given by
\begin{eqnarray}
  \label{eq:fd_masses}
M_H^2 \! &=&\! -2\lambda_{345}\left(\Phi_1^{\dagger} \Phi_1 
               +\Phi_2^{\dagger}\Phi_2\right)\;,\qquad\   
M_A^2\ =\ -2\lambda_{5}\left(\Phi_1^{\dagger} \Phi_1 
+ \Phi_2^{\dagger} \Phi_2\right)\; ,\nonumber \\[2mm]
M_{H^\pm}^2 \!&=&\! -\lambda_{45}\left(\Phi_1^{\dagger} \Phi_1
                           +\Phi_2^{\dagger} \Phi_2\right)\;,
\qquad\quad M_W^2 \ =\ \frac{g^2}{2}\left(\Phi_1^{\dagger} \Phi_1
                     +\Phi_2^{\dagger} \Phi_2\right)\;, \nonumber\\[2mm] 
M_Z^2 \!&=&\! \frac{g^2}{2c_w^2}\left(\Phi_1^{\dagger} \Phi_1
      +\Phi_2^{\dagger} \Phi_2\right)\;,\qquad\qquad
m_t^2 \ =\ |h_I|^2 \Phi_I^{\dagger} \Phi_I\; .
\end{eqnarray}
In   the    above,   we    have   used   the    short-hand   notation:
$\lambda_{ij(k)}=\lambda_{i}+\lambda_{j}(+\lambda_{k})$,           with
$i,j,k=3,4,5$,  and labelled  with $I=1$  or $I=2$,  according  to the
${\rm Z}_2$ symmetry.

Adding  the  one-loop  effective  potential  to  the  tree-level  one,
i.e.~$V=V^0+V^{\rm 1-loop}_{\rm eff}$, the tadpole conditions now read:
\begin{eqnarray}
\bigg<\frac{\partial V}{\partial \phi_1}\bigg>\ =\
T_{\phi_1} +
\bigg<\frac{\partial V^{\rm 1-loop}_{\rm eff}}{\partial \phi_1}\bigg>
\ =\ 0\; , \qquad
\bigg<\frac{\partial V}{\partial \phi_2}\bigg>\ =\
T_{\phi_2} +
\bigg<\frac{\partial V^{\rm 1-loop}_{\rm eff}}{\partial \phi_2}\bigg>
\ =\ 0\; .
\end{eqnarray}
More explicitly, we obtain
\begin{eqnarray}
\bigg<\frac{\partial V^{\rm 1-loop}_{\rm eff}}{\partial \phi_i}\bigg>\
=\ \frac{v_i\,v^2}{64\pi^2}\,\Delta\widehat{t}_i\; ,
\end{eqnarray}
where $\Delta\widehat{t}_{1,2}$ are found to be
\begin{eqnarray}
\Delta\widehat{t}_i \!&=&\! \frac{1}{v^2}\,\bigg[
 4\lambda_{345} M_H^2 \left(1-\ln\frac{M_H^2}{Q^2}\right)
+4\lambda_{5} M_A^2 \left(1-\ln\frac{M_A^2}{Q^2}\right)
+4\lambda_{45} M_{H^\pm}^2 \left(1-\ln\frac{M_{H^\pm}^2}{Q^2}\right)
\nonumber \\[3mm] &&\hspace{-1.5cm}
-6g^2 M_W^2 \left(\frac{1}{3}-\ln\frac{M_W^2}{Q^2}\right)
-3\frac{g^2}{c_W^2} M_Z^2
\left(\frac{1}{3}-\ln\frac{M_Z^2}{Q^2}\right)
+12|h_I|^2 m_t^2 \left(1-2\ln\frac{m_t^2}{Q^2}\right)\,\delta_{Ii}
\bigg] .\qquad
\end{eqnarray}
Thus, the one-loop improved tadpole conditions are given by
\begin{eqnarray}
  \label{eq:1looptad}
\frac{T_{\phi_1}}{vc_\beta}\: +\:
\frac{v^2\,\Delta\widehat{t}_1}{64\pi^2}\  =\ 0\;,  
\qquad
\frac{T_{\phi_2}}{vs_\beta}\: +\:
\frac{v^2\,\Delta\widehat{t}_2}{64\pi^2}\ =\ 0\; . 
\end{eqnarray}
These conditions  can easily be solved for  the quartic
couplings $\lambda_{1}$
and     $\lambda_{2}$,    in    terms     of    the     other    three
couplings~$\lambda_{3,4,5}$.

\subsubsection{Masses of the CP-odd neutral and charged Higgs bosons}

The  one-loop corrected  potential  term for  the  CP-odd scalar  mass
matrix reads:
\begin{equation}
V_{\rm mass}^{\rm CP-odd}\ =\ \frac{1}{2}\,
\Big( a_1 \,,\ a_2 \Big)\,{\cal M}^2_P\,
\left(\begin{array}{c} a_1 \\ a_2 \end{array} \right)\;,
\end{equation}
where
\begin{equation}
{\cal M}^2_P\ =\ \left(\begin{array}{cc}
-\lambda_5 v^2 s_\beta^2 +\frac{T_{\phi_1}}{vc_\beta}+
\left\langle\frac{\partial^2 V^{\rm 1-loop}_{\rm eff}}{\partial a_1^2}\right\rangle
&
\lambda_{5} v^2 c_\beta s_\beta +
\left\langle\frac{\partial^2 V^{\rm 1-loop}_{\rm eff}}{\partial a_1\partial a_2}
\right\rangle
\\[0.3cm]
\lambda_{5} v^2 c_\beta s_\beta +
\left\langle\frac{\partial^2 V^{\rm 1-loop}_{\rm eff}}{\partial a_1\partial a_2}
\right\rangle
&
-\lambda_5 v^2 c_\beta^2 +\frac{T_{\phi_2}}{vs_\beta}+
\left\langle
\frac{\partial^2 V^{\rm 1-loop}_{\rm eff}}{\partial a_2^2}
\right\rangle
\end{array}\right)\,. \\[2mm]
\end{equation}
The VEVs of the double derivatives are found to be
\begin{equation}
  \label{eq:dderiv}
\bigg<\frac{\partial^2 V^{\rm 1-loop}_{\rm eff}}{\partial a_i\partial a_j}
\bigg>\ =\ \frac{v^2}{64\pi^2}\Delta\widehat{t}_i\,\delta_{ij}\; .
\end{equation}
Employing  the one-loop tadpole  conditions~(\ref{eq:1looptad}), along
with~(\ref{eq:dderiv}), we  find that  the CP-odd mass  matrix retains
its tree-level form, i.e.
\begin{equation}
{\cal M}^2_P\  =\ M_A^2\,\left(\begin{array}{cc}
s_\beta^2 & -c_\beta s_\beta \\[0.3cm]
-c_\beta s_\beta & c_\beta^2
\end{array}\right)
\end{equation}
with $M_A=-\lambda_5v^2$.  In similar  fashion, we have that radiative
effects  do  not  modify  the  tree-level  structure  of  the  charged
Higgs-boson mass matrix:
\begin{equation}
V_{\rm mass}^{H^\pm}\ =\ M_{H^\pm}^2\,
\Big( \phi^-_1 \,,\ \phi^-_2 \Big)\,\left(\begin{array}{cc}
s_\beta^2 & -c_\beta s_\beta \\[0.3cm]
-c_\beta s_\beta & c_\beta^2
\end{array}\right)\,
\left(\begin{array}{c} \phi^+_1 \\ \phi^+_2 \end{array} \right)\, ,
\end{equation}
with $M_{H^\pm}^2=-\lambda_{45}\,v^2/2$.

\subsubsection{Masses and mixing of the CP-even neutral Higgs bosons}

One-loop quantum effects give rise to non-trivial contributions to the
masses  of the  CP-even neutral  Higgs  bosons and  their mixing.  The
one-loop-corrected potential term  describing these quantum effects is
given by
\begin{equation}
V_{\rm mass}^{\rm CP-even}\ =\ \frac{1}{2}\,
\Big( \phi_1 \,,\ \phi_2 \Big)\,{\cal M}^2_S\,
\left(\begin{array}{c} \phi_1 \\ \phi_2 \end{array} \right)\; ,
\end{equation}
where ${\cal M}^2_S$ is the $2\times 2$ one-loop improved CP-even mass
matrix
\begin{equation}
{\cal M}^2_S\  =\ \left(\begin{array}{cc}
2\lambda_1 v^2 c_\beta^2 +\frac{T_{\phi_1}}{vc_\beta}+
\left\langle\frac{\partial^2 V^{\rm 1-loop}_{\rm eff}}{\partial
  \phi_1^2}\right\rangle 
&
\lambda_{345} v^2 c_\beta s_\beta +
\left\langle\frac{\partial^2 V^{\rm 1-loop}_{\rm eff}}{\partial
  \phi_1\partial \phi_2} 
\right\rangle
\\[0.3cm]
\lambda_{345} v^2 c_\beta s_\beta +
\left\langle\frac{\partial^2 V^{\rm 1-loop}_{\rm eff}}{\partial
  \phi_1\partial \phi_2} 
\right\rangle
&
2\lambda_2 v^2 s_\beta^2 +\frac{T_{\phi_2}}{vs_\beta}+
\left\langle
\frac{\partial^2 V^{\rm 1-loop}_{\rm eff}}{\partial \phi_2^2}
\right\rangle
\end{array}\right)\,.
\end{equation}
Here, the  VEVs of the  double derivatives of the  effective potential
with respect to the  CP-even scalar fields $\phi_{1,2}$ are calculated
to be
\begin{equation}
\bigg<\frac{\partial^2 V^{\rm 1-loop}_{\rm eff}}{\partial
  \phi_i\partial \phi_j} 
\bigg>\ =\ \frac{1}{64\pi^2}\left(
v_iv_j \Delta\widehat{m}^2_{ij}+v^2\Delta\widehat{t}_i \delta_{ij}
\right)\; ,
\end{equation}
with
\begin{eqnarray}
  \label{eq:Dmhat2ij}
\Delta\widehat{m}^2_{ij} &\equiv&
 8\lambda_{345}^2\ln\frac{|M_H|^2}{Q^2}
+8\lambda_{5}^2\ln\frac{M_A^2}{Q^2}
+4\lambda_{45}^2\ln\frac{M_{H^\pm}^2}{^2}
+g^4\left(2+3\ln\frac{M_W^2}{Q^2}\right)\nonumber\\
&&+\: \frac{g^4}{2c_W^4}\left(2+3\ln\frac{M_Z^2}{Q^2}\right)
-12 |h_I|^4
\left(1+2\ln\frac{m_t^2}{Q^2}\right)\,\delta_{ij}\,\delta_{Ii}\; .
\end{eqnarray}
After implementing the one-loop tadpole conditions~(\ref{eq:1looptad}),
the CP-even scalar mass matrix ${\cal M}^2_S $ simplifies to
\begin{equation}
{\cal M}^2_S\  =\ v^2\,\left(\begin{array}{cc}
\left(2\lambda_1+\frac{\Delta\widehat{m}^2_{11}}{64\pi^2}\right)  c_\beta^2
&
\left(\lambda_{345}+\frac{\Delta\widehat{m}^2_{12}}{64\pi^2}\right)
c_\beta s_\beta 
\\[0.3cm]
\left(\lambda_{345}+\frac{\Delta\widehat{m}^2_{12}}{64\pi^2}\right)
c_\beta s_\beta 
&
\left(2\lambda_2+\frac{\Delta\widehat{m}^2_{22}}{64\pi^2}\right)  s_\beta^2
\end{array}\right)\,.
\end{equation}
Notice that  the top-quark contribution  in~(\ref{eq:Dmhat2ij}) breaks
the universality of $\Delta\widehat{m}^2_{ij}$.

In contrast to what happens  at the tree-level, the diagonalization of
the  one-loop   effective  mass  matrix  ${\cal   M}^2_S$  yields  two
non-vanishing mass  eigenvalues. As a  consequence of the  breaking of
the scaling symmetry at  the quantum level, the pseudo-Goldstone boson
$h$ receives  a radiative  mass, which could  be even larger  than the
non-zero tree-level  mass $M_H$,  for specific choices  of parameters.
To appropriately describe the  radiatively corrected masses and mixing
of the  CP-even Higgs  bosons, we introduce  a $2\times  2$ orthogonal
matrix~$O$, through
\begin{equation}
\big(\phi_1\,,\ \phi_2\big)^{\sf T}_\alpha \ = \
O_{\alpha i}\; \big(H_1\,,\ H_2\big)^{\sf T}_i\; ,
\end{equation}
which  diagonalizes  the  CP-even  mass  matrix as  $O^{\sf  T}  {\cal
  M}_S^2\,O  \  =   \  {\rm  diag}(M_{H_1}^2\,,M_{H_2}^2)$,  with  the
convention $M_{H_1} \leq M_{H_2}$.

In terms of  the mixing matrix $O$, the couplings  of the Higgs bosons
to the vector bosons are given by
\begin{eqnarray}
  \label{eq:LHVV}
{\cal L}_{HVV} \!& = &\! g\,M_W \, 
\sum_i \,g_{H_iVV}\,  \left( H_i\, W^+_\mu W^{- \mu}\: + \: 
\frac{1}{2c_W^2}\, H_i\, Z_\mu Z^\mu\right) \, ,\\[3mm]
  \label{eq:LHAZ}
{\cal L}_{HAZ} \!&=&\! \frac{g}{2c_W} \sum_{i} g_{H_iAZ}\, Z^{\mu}
(A\, \stackrel {\leftrightarrow} {\partial}_\mu H_i) \,, \\ [3mm]
  \label{eq:LHplus}
{\cal L}_{HH^\pm W^\mp} \!&=&\! \frac{g}{2} \,\left[ \sum_i \, 
g_{H_iH^- W^+}\, 
W^{+\mu} (H_i\, i\!\stackrel{\leftrightarrow}{\partial}_\mu H^-)\, 
+\, W^{+\mu} (A\, i\!\stackrel{\leftrightarrow}{\partial}_\mu H^-)\, 
+\, {\rm h.c.}\,\right] ,\quad
\end{eqnarray}
where the action  of $\stackrel{\leftrightarrow}{\partial}_\mu$ on two
arbitrary  functions $f(x)$  and $g(x)$  is defined,  such  that $f(x)
\stackrel{\leftrightarrow}{\partial}_\mu      g(x)     \equiv     f(x)
(\partial_\mu  g(x)) -  (\partial_\mu f(x))  g(x)$.  In  addition, the
reduced couplings that occur in~(\ref{eq:LHVV})--(\ref{eq:LHplus}) are
given by
\begin{eqnarray}
g_{H_iVV}\ =\ c_\beta\,O_{1i}+s_\beta\,O_{2i} \; ,\qquad
g_{H_iAZ}\ =\ g_{H_iH^- W^+}\ =\  c_\beta\,O_{2i}-s_\beta\,O_{1i} \; ,
\end{eqnarray}
which satisfy the identity
\begin{equation}
g_{H_iAZ}^2\: +\: g_{H_iVV}^2 \ = \ 1\; ,
\end{equation}
for  each  $i  =  1,2$.    The  latter  implies  that  $g_{H_1AZ}^2  =
g_{H_2VV}^2$ and $g_{H_2AZ}^2 = g_{H_1VV}^2$.

\begin{figure}[!t]
\begin{center}
{\epsfig{figure=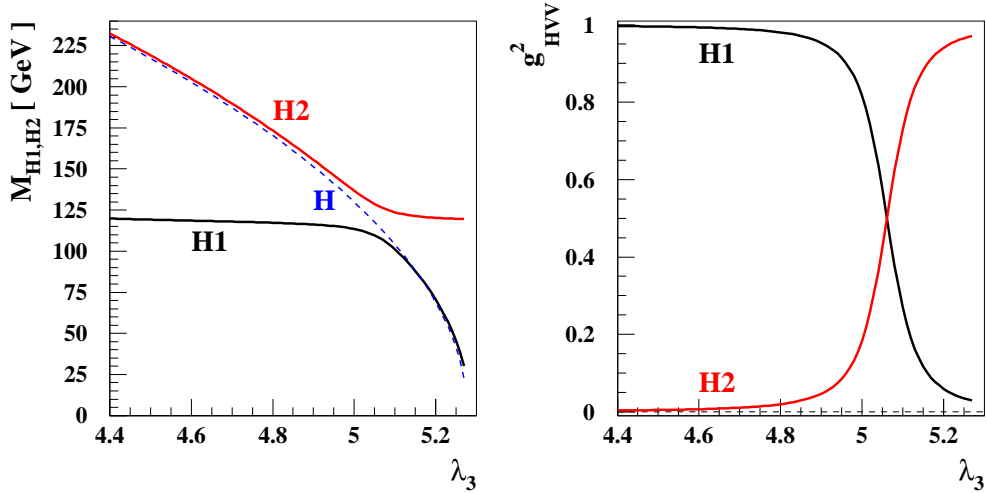,height=15.0cm,width=15.0cm}}
\end{center}
\vspace{-8.0cm}
\caption{\it The CP-even Higgs masses (left panel) and their couplings
$g_{H_iVV}^2$ (right panel), as  functions of $\lambda_3$. We have set
$\tan\beta=1$    and    $M_A=M_{H^\pm}=400$~GeV,   corresponding    to
$\lambda_4=\lambda_5\simeq     -2.64$.     The     parameter     $M_H=
\sqrt{-\lambda_{345}}\, v$ is the tree-level CP-even Higgs-boson mass.
The RG scale  $Q = \Lambda_{\rm GW}$ is chosen; see  the text for more
details.  }
\label{fig:mhghsq}
\end{figure}

For  illustration,   we  show  in   Figure~\ref{fig:mhghsq}  numerical
estimates of the CP-even Higgs-boson masses $M_{H_1,H_2}$ (left panel)
and  their  couplings $g_{H_iVV}^2$  (right  panel),  as functions  of
$\lambda_3$.   We have taken  $\tan\beta=1$ and  fixed the  CP-odd and
charged  Higgs-boson  masses   to  be:  $M_A=M_{H^\pm}=400$~GeV.   The
dependence     of     the     tree-level    CP-even     Higgs     mass
$M_H=\sqrt{-\lambda_{345}}\, v$ on  $\lambda_3$ is also displayed with
a dashed line.   We observe that there is  a level-crossing phenomenon
taking  place at  the critical  value $\lambda_3  =  \lambda^{\rm c}_3
\simeq  5.06$,   at  which  $g_{H_1VV}^2=g_{H_2VV}^2$.    For  quartic
couplings  $\lambda_3$ smaller than  $\lambda^{\rm c}_3$,  the lighter
state $H_1$ is  mainly SM-like and has the larger  coupling to the $Z$
boson,  i.e.~$g_{H_1VV}^2 >  g_{H_2VV}^2$, whereas  the  heavier boson
$H_2$  has a smaller  coupling to  $Z$ and  its mass  is close  to the
tree-level value, i.e.~$M_{H_2}\sim M_H$. If $\lambda_3 > \lambda^{\rm
c}_3$, the  roles of the $H_1$  and $H_2$ bosons  get exchanged, where
the  heavier  state  $H_2$  becomes  the  SM-like  Higgs  boson,  with
$g_{H_2VV}^2 > g_{H_1VV}^2$, and $M_{H_1}\sim M_H$.

Before closing this section, we comment on our choice of the RG scale:
\begin{equation}
  \label{eq:QLGW}
Q\ =\ \Lambda_{\rm GW}\; ,
\end{equation}
where   $\Lambda_{\rm   GW}$   is  the   so-called   Gildener-Weinberg
scale~\cite{Gildener:1976ih}   which  may   be  determined   from  the
expression
\begin{equation}
\ln\frac{\Lambda_{\rm GW}}{v}\ =\ \frac{\cal A}{2{\cal B}}\:
+\: \frac{1}{4}\ .
\end{equation}
Here, the parameters ${\cal A}$ and ${\cal B}$ are given by
\begin{eqnarray}
  \label{eq:calab}
{\cal A}&=&\frac{1}{64\pi^2 v^4}\left[
 M_H^4\left(-\frac{3}{2}+\ln\frac{M_H^2}{v^2}\right)
+M_A^4\left(-\frac{3}{2}+\ln\frac{M_A^2}{v^2}\right)
+2M_{H^\pm}^4\left(-\frac{3}{2}+\ln\frac{M_{H^\pm}^2}{v^2}\right)\right.
\nonumber \\ && \left.
+6M_W^4\left(-\frac{5}{6}+\ln\frac{M_W^2}{v^2}\right)
+3M_Z^4\left(-\frac{5}{6}+\ln\frac{M_Z^2}{v^2}\right)
-12m_t^4\left(-1+\ln\frac{m_t^2}{v^2}\right)
\right]\,,\nonumber\\[3mm]
{\cal B}&=&\frac{1}{64\pi^2 v^4}\left(
M_H^4 + M_A^4 + 2 M_{H^\pm}^4 +6 M_W^4 + 3 M_Z^4 -12 m_t^4 \right)\,.
\label{eq:calB}
\end{eqnarray}
With the choice for the RG scale $Q$ given in~(\ref{eq:QLGW}), we have
checked  that   the  radiative  corrections  are   minimized  and  the
predictions for  the masses  of the CP-even  Higgs bosons  exhibit the
least sensitivity, under small  variations of $Q$ around $\Lambda_{\rm
  GW}$.   We note  that in  kinematic  regions far  from the  critical
level-crossing point, e.g.~for  $\lambda_3 \ll \lambda^{\rm c}_3$, the
tree-level relations $M_{H_2}\simeq  M_H$, $g_{H_2VV}^2 \simeq 0$, and
$g_{H_1VV}^2  \simeq  1$  prove  to  be  an  excellent  approximation.
Moreover, the  radiative mass $M_{H_1}$ of  the pseudo-Goldstone boson
$H_1$ may  well be approximated by the  Gildener-Weinberg mass $M_{\rm
  GW}$:
\begin{equation}
  \label{eq:mgw}
M_{H_1}^2\ \simeq\ M_{\rm GW}^2\ \equiv\ 8 {\cal B} v^2\; ,
\end{equation}
where the parameter ${\cal B}$ is given by~(\ref{eq:calab}).

\section{Numerical Analysis}
\label{sec:analysis}

The  SI-2HDM  may  be  parameterized,  in terms  of  five  independent
kinematic  parameters.   These parameters  could  be  either the  five
quartic                                                       couplings
$(\lambda_1,\lambda_2,\lambda_3,\lambda_4,\lambda_5)$,   or   the  set
$(v,t_\beta,M_H,M_{H^\pm},M_A)$. At  the tree-level, the  two sets are
simply related, by  means of~(\ref{eq:set2}) and~(\ref{eq:set1}).  For
our numerical analysis, we choose to vary the four parameters:
\begin{eqnarray}
t_\beta \;,\quad M_{H^\pm}\;,\quad  M_A\;,\quad M_H^{\rm eff}\;,
\end{eqnarray}
with $v \simeq 246$~GeV and
\begin{equation}
M_H^{\rm eff}\ \equiv\ M_{H_2} g_{_{H_1VV}}^2 + M_{H_1} g_{_{H_2VV}}^2\; .
\end{equation}
The latter mass parameter was  introduced, since its value stays close
to the  one of  the tree-level $H$-boson  mass $M_H$,  after radiative
corrections are  included. As discussed  in the previous  section, the
masses  of the charged  and CP-odd  Higgs bosons  are not  affected by
quantum  effects, so  the  couplings $\lambda_4$  and $\lambda_5$  are
determined  by  the  tree-level  relations  given  in~(\ref{eq:set1}).
Instead, the  couplings $\lambda_{1,2,3}$ receive  significant quantum
corrections beyond the Born approximation. Explicitly, for given input
values   of    $M_H^{\rm   eff}$   and    $t_\beta$,   the   couplings
$\lambda_{1,2,3}$  can be  determined iteratively,  after  taking into
consideration the  one-loop tadpole conditions in~(\ref{eq:1looptad}).
For definiteness,  we have assumed  the Type-II Yukawa sector  for the
top-quark      mass       $m_t$,      corresponding      to      $I=2$
in~(\ref{eq:fd_masses}). However,  our results  do not depend  on this
choice.

\subsection{Theoretical and Phenomenological Constraints}

We now  consider several theoretical  and phenomenological constraints
on  the  SI-2HDM.    These  include:  (i)~the  perturbative  unitarity
bounds~\cite{Lee:1977yc,Lee:1977eg},  (ii)  the  indirect  constraints
from the electroweak precision data~\cite{Nakamura:2010zzi}, and (iii)
the direct constraints  from the LEP collider~\cite{Schael:2006cr} and
the LHC~\cite{LHC_SM_Higgs}.

We first  consider the constraints  obtained by requiring  validity of
perturbative    unitarity~\cite{Lee:1977yc,Lee:1977eg}.     For    the
tree-level        unitarity        conditions,       we        closely
follow~\cite{Kanemura:1993hm}.   We   observe  that  the  perturbative
unitarity  constraint  is  weakest,  when $\tan\beta=1$,  and  becomes
stronger, as $\tan\beta$ deviates from  this value. The reason is that
the  couplings $\lambda_1  \propto t_\beta^2$  and  $\lambda_2 \propto
1/t_\beta^2$   for   the   present  ${\rm   Z}_2$-invariant   SI-2HDM.
Furthermore, at the tree  level, the perturbative unitarity bounds are
symmetric under the  exchange $c_\beta \leftrightarrow s_\beta$, since
the eigenvalues of the  scattering matrices depend on the combinations
of $\lambda_1 + \lambda_2$  and $(\lambda_1 - \lambda_2)^2$, while the
other  couplings  $\lambda_{3,4,5}$  are independent  of  $\tan\beta$.
Specifically,  one of  the  most stringent  conditions  may come  from
requiring    that   the   eigenvalue    $a_+$   of    the   scattering
matrices~\cite{Kanemura:1993hm} obeys the bound:
\begin{equation}
a_+\ \equiv\ 
\frac{1}{16\pi}\left[3(\lambda_1+\lambda_2)
+\sqrt{9(\lambda_1-\lambda_2)^2+(2\lambda_3+\lambda_4)^2}\right] 
\ \leq\ \frac{1}{2}\ .
\end{equation}
In view of the above discussion, we only consider regions of parameter
space, for which $\tan\beta \geq 1$.

The electroweak oblique corrections to  the so-called $S$, $T$ and $U$
parameters~\cite{Peskin:1990zt,Peskin:1991sw}    provide   significant
constraints on the quartic couplings  of the SI-2HDM.  For a vanishing
$U$  parameter   ($U=0$),  the  electroweak   oblique  parameters  are
constrained by the following inequality:
\begin{equation}
\frac{(S-\widehat S_0)^2}{\sigma_S^2}\ +\
\frac{(T-\widehat T_0)^2}{\sigma_T^2}\ -\
2\rho_{ST}\frac{(S-\widehat S_0)(T-\widehat T_0)}{\sigma_S \sigma_T}\
\leq\ R^2\,(1-\rho_{ST}^2)\; ,
\end{equation}
with $R^2=2.30$, $4,61$, $5.99$  and $9.21$, for electroweak precision
limits at  $68 \%$,  $90 \%$,  $95 \%$ and  $99 \%$  confidence levels
(CLs), respectively.  The central values and their standard deviations
are given by~\cite{Nakamura:2010zzi}
\begin{equation}
(\widehat S_0\,,\ \sigma_S)\ =\ (0.03\,,\ 0.09)\;,\qquad
(\widehat T_0\,,\ \sigma_T)\ =\ (0.07 \,,\ 0.08)\; ,   
\end{equation}
for the  value $\rho_{ST}=0.82$ of the correlation  parameter.  In our
numerical analysis, we apply the 90\% CL limits.

The SI-2HDM  contributions~\cite{Toussaint:1978zm} to the  $S$ and $T$
parameters may conveniently be expressed as follows:
\begin{eqnarray}
  \label{eq:STphi}
S_\Phi \!&=&\! -\frac{1}{4\pi} \left[
\left(1+\delta_{\gamma Z}^{H^\pm}\right)^2F^\prime_\Delta(M_{H^\pm},M_{H^\pm})
-\sum_{i=1,2}\left(g_{H_iAZ}+\delta_Z^{H_i}\right)^2 F^\prime_\Delta(M_{H_i},M_A)
\right]\,, \nonumber\\
T_\Phi \!&=&\! -\frac{\sqrt{2}G_F}{16\pi^2\alpha_{\rm EM}}\ \Bigg\{
-\left(1+\delta_W^{A}\right)^2F_\Delta(M_A,M_{H^\pm})\\
&&  
+\sum_{i=1,2}\left[
\left(g_{H_iAZ}+\delta_Z^{H_i}\right)^2 F_\Delta(M_{H_i},M_A)
-\left(g_{H_iH^- W^+}+\delta_W^{H_i}\right)^2 F_\Delta(M_{H_i},M_{H^\pm})
\right] \Bigg\}\; .\quad   \nonumber
\end{eqnarray}
In the evaluation of  the new-physics parameters $S_\Phi$ and $T_\Phi$
in  (\ref{eq:STphi}), we have  dressed the  vertex couplings  with the
dominant             one-loop            corrections            ${\cal
O}\left(\lambda^2/16\pi^2\right)$,   where   $\lambda$  symbolizes   a
generic  quartic  coupling  $\lambda_{1,2,3,4,5}$.   These  additional
$\lambda^2$-dependent  contributions  are  denoted as  $\delta_{\gamma
Z}^{H^\pm}$, $\delta_{Z}^{H_1,H_2}$  and $\delta_{W}^{A,H_1,H_2}$, and
become  rather  significant for  quartic  couplings  $|\lambda| >  1$.
Their explicit analytic forms are presented in Appendix~\ref{sec:vtx}.

On  the  other hand,  the  analytic  form  of the  one-loop  functions
$F_\Delta(m_1,m_2)$   and  $F^\prime_\Delta(m_1,m_2)$  may   be  found
in~\cite{Kanemura:2011sj}.   Here we  simply quote  some of  their key
properties:                      $F_\Delta(m_1,m_2)=F_\Delta(m_2,m_1)$,
$F^\prime_\Delta(m_1,m_2)=F^\prime_\Delta(m_2,m_1)$                 and
$F_\Delta(m,m)=0$.   If the  $\lambda^2$-dependent  vertex corrections
are  ignored,  then  $S_\Phi$   and  $T_\Phi$  become  independent  of
$\tan\beta$  and  symmetric under  the  exchange $M_A  \leftrightarrow
M_{H_2}$,   since  $g_{H_2AZ}^2=g_{H_2H^-   W^+}^2=g_{H_1VV}^2=1$  and
$g_{H_1AZ}^2=g_{H_1H^- W^+}^2=g_{H_2VV}^2=0$ at  the tree level in the
SI-2HDM.  Finally, it is interesting to observe that $T_\Phi$ vanishes
identically, in  the limit $M_A  \to M_{H^\pm}$, or  equivalently when
$\lambda_4  \to \lambda_5$.  In  this limit,  the SI-2HDM  realizes an
unbroken SO(3)  custodial symmetry in the bilinear  scalar field space
of  SO(5),   according  to  a   recent  classification  of   the  2HDM
potential~\cite{Battye:2011jj,Pilaftsis:2011ed}.   Since this symmetry
remains unbroken  even by the inclusion  of $\lambda$-dependent vertex
corrections, the electroweak parameter $T_\Phi$ still vanishes.

The total  contribution to the  electroweak $S$ and $T$  parameters is
given  by   the  sums:  $S=S_{\rm  SM}+S_{\rm   \Phi}$  and  $T=T_{\rm
  SM}+T_{\rm \Phi}$.   For the SM  contribution, we have  employed the
parameterizations~\cite{Cho:1999km}:
\begin{eqnarray}
S_{\rm SM} \!&=&\! -0.007 x_t + 0.091 x_h - 0.010 x_h^2\;,  \nonumber \\
T_{\rm SM} &=& (0.130-0.003 x_h) x_t + 0.003 x_t^2
-0.079 x_h -0.028 x_h^2 + 0.0026 x_h^3\; ,
\end{eqnarray}
with    $x_t=(m_t/{\rm   GeV}    -173)/10$    and   $x_h=\ln(M_{H_{\rm
    SM}}/117~{\rm     GeV})$,      where     $M_{H_{\rm     SM}}\equiv
M_{H_1}g_{H_1VV}^2+M_{H_2}g_{H_2VV}^2$.     This    last    expression
approximates the mass of the SM Higgs boson fairly well over the whole
region of the parameter space.

The  recent LHC  data  pertinent to  SM  Higgs-boson searches  provide
important constraints  on the kinematic parameters of  the SI-2HDM. In
our numerical  analysis, we derive conservative limits  by taking that
either $g_{H_1VV}^2=1$, or $g_{H_2VV}^2=1$.   To this end, we consider
the  95\%~CL exclusion limits  on the  SM Higgs-boson  mass $M_{H_{\rm
    SM}}$,     as     quoted      by     the     CMS     and     ATLAS
collaborations~\cite{LHC_SM_Higgs}:
\begin{eqnarray}
{\rm CMS~:} &&  ~~127 \ {\rm GeV} -\ 600 \ {\rm GeV}\;, 
\nonumber \\
{\rm ATLAS~:} &&  112.7 \ {\rm GeV} - 115.5 \ {\rm GeV}\;, \qquad
131 \ {\rm GeV} - 453 \ {\rm GeV}\,.
\end{eqnarray}
Combining the above CMS and ATLAS results, the following LHC exclusion
limits on the Higgs masses may be deduced:
\begin{eqnarray}
&&\hspace{-2.0cm}
127\ <\ M_{H_1}/{\rm GeV}\ <\ 600\;, 
\quad \mbox{when} ~~ g_{H_1VV}^2 \geq 0.99\;, \nonumber \\
&&\hspace{-2.0cm}
127\ <\ M_{H_2}/{\rm GeV}\ <\ 600\;,\quad  
\mbox{when} ~~ g_{H_2VV}^2 \geq 0.99\;.  \nonumber
\end{eqnarray}
More precise limits may be derived by calculating the production cross
sections for each Higgs search  channel, in conjunction with the limits
on the  ratio $\sigma/\sigma_{\rm  SM}$.  We leave  this issue  to our
experimental colleagues for more  detailed analyses.  Finally, we have
included the LEP limits according to~\cite{Schael:2006cr}.

\subsection{Numerical Predictions}

\begin{figure}[!t]
\begin{center}
{\epsfig{figure=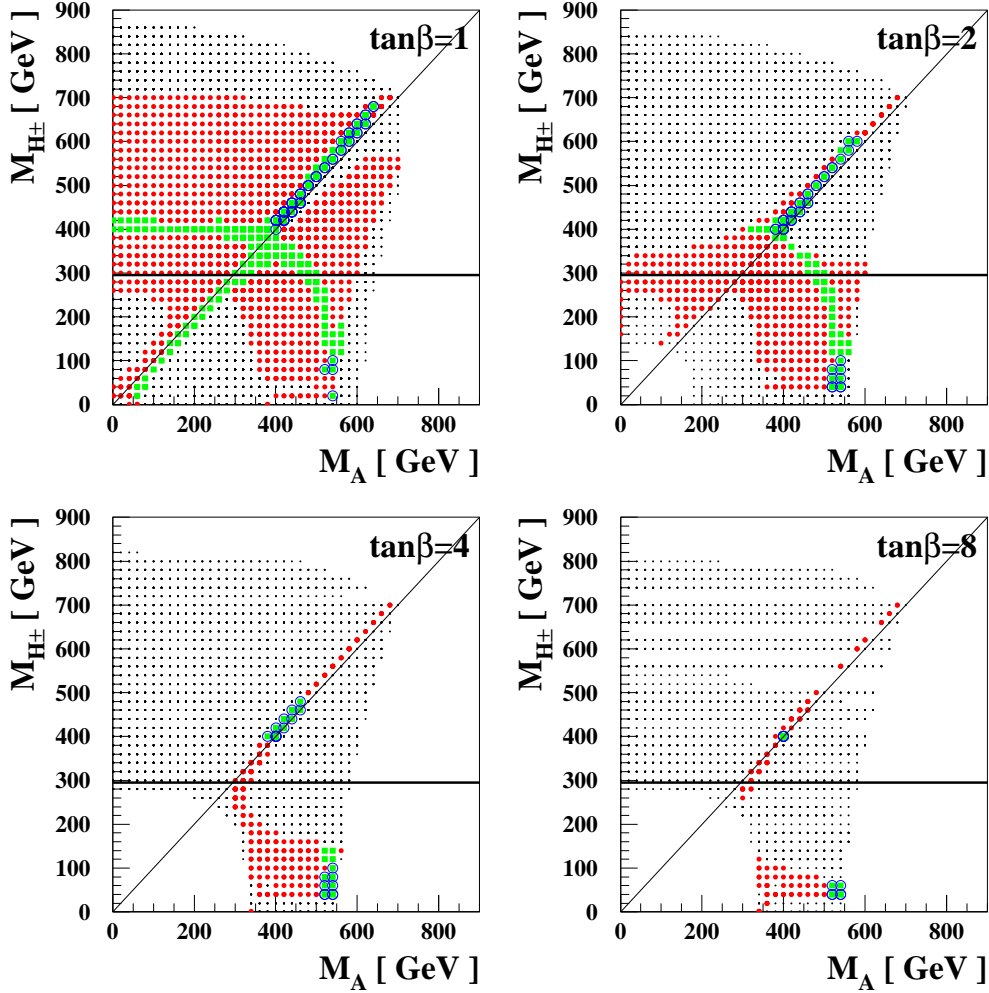,height=15.0cm,width=15.0cm}}
\end{center}
\vspace{-1.0cm}
\caption{\it The allowed parameter  space in the $M_A-M_{H^\pm}$ plane
  compatible  with  perturbative  unitarity  (black)  and  electroweak
  precision limits  (red) at the 90\%  CL, for $\tan\beta  = 1$ (upper
  left panel),  $\tan\beta = 2$  (upper right panel), $\tan\beta  = 4$
  (lower left  panel) and  $\tan\beta = 8$  (lower right  panel).  The
  green  region indicates  the allowed  area due  to the  LEP  and LHC
  limits.  The  little blue circles on  the green area  single out the
  region, for which $|g_{H_2VV}| > |g_{H_1VV}|$.  The thick horizontal
  line   gives   a   lower   bound   on   the   charged   Higgs   mass
  $M_{H^\pm}\protect\gsim   295$   GeV,   from  the   $b\to   s\gamma$
  data~\cite{Misiak:2006zs}, assuming Type-II Yukawa couplings.  }
\label{fig:mch_ma}
\end{figure}
\begin{figure}[!t]
\begin{center}
{\epsfig{figure=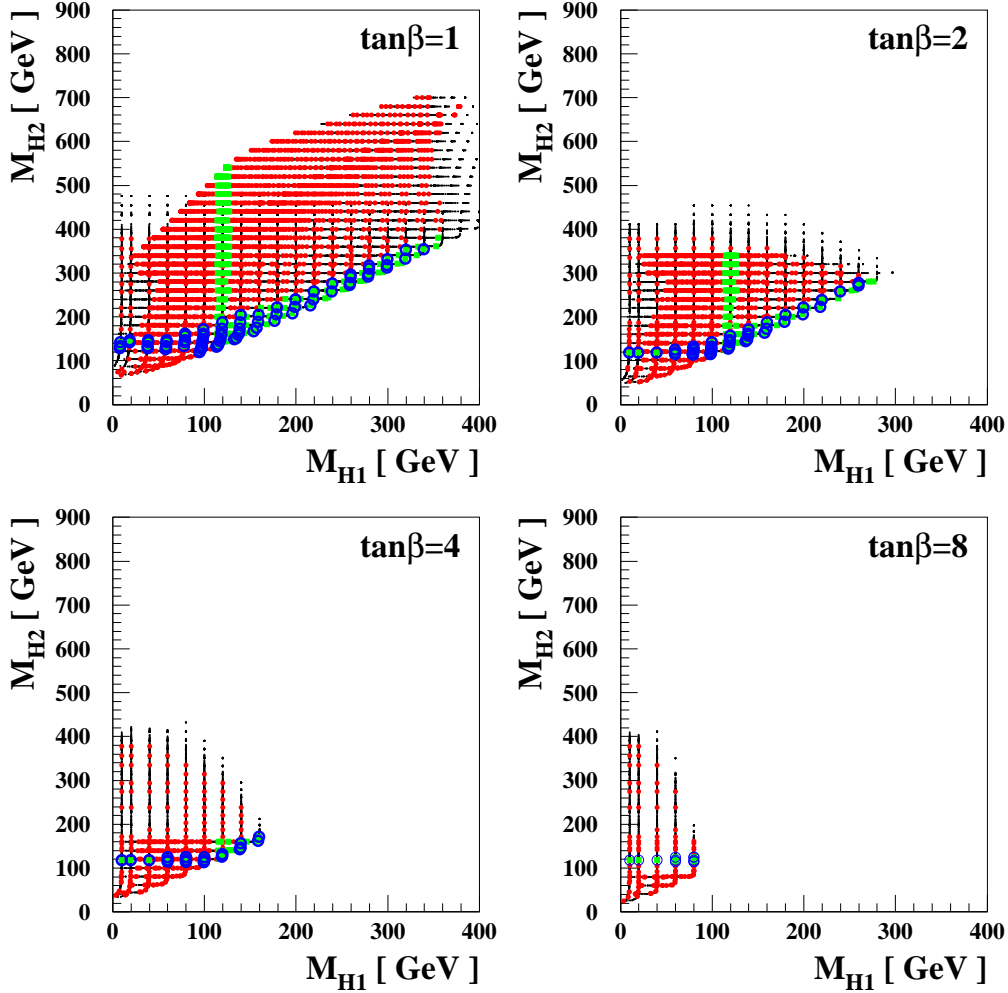,height=15.0cm,width=15.0cm}}
\end{center}
\vspace{-1.0cm}
\caption{\it The same as in Fig.~\ref{fig:mch_ma},
but in the $M_{H1}$-$M_{H_2}$ plane
}
\label{fig:mh12}
\end{figure}
\begin{figure}[!t]
\begin{center}
{\epsfig{figure=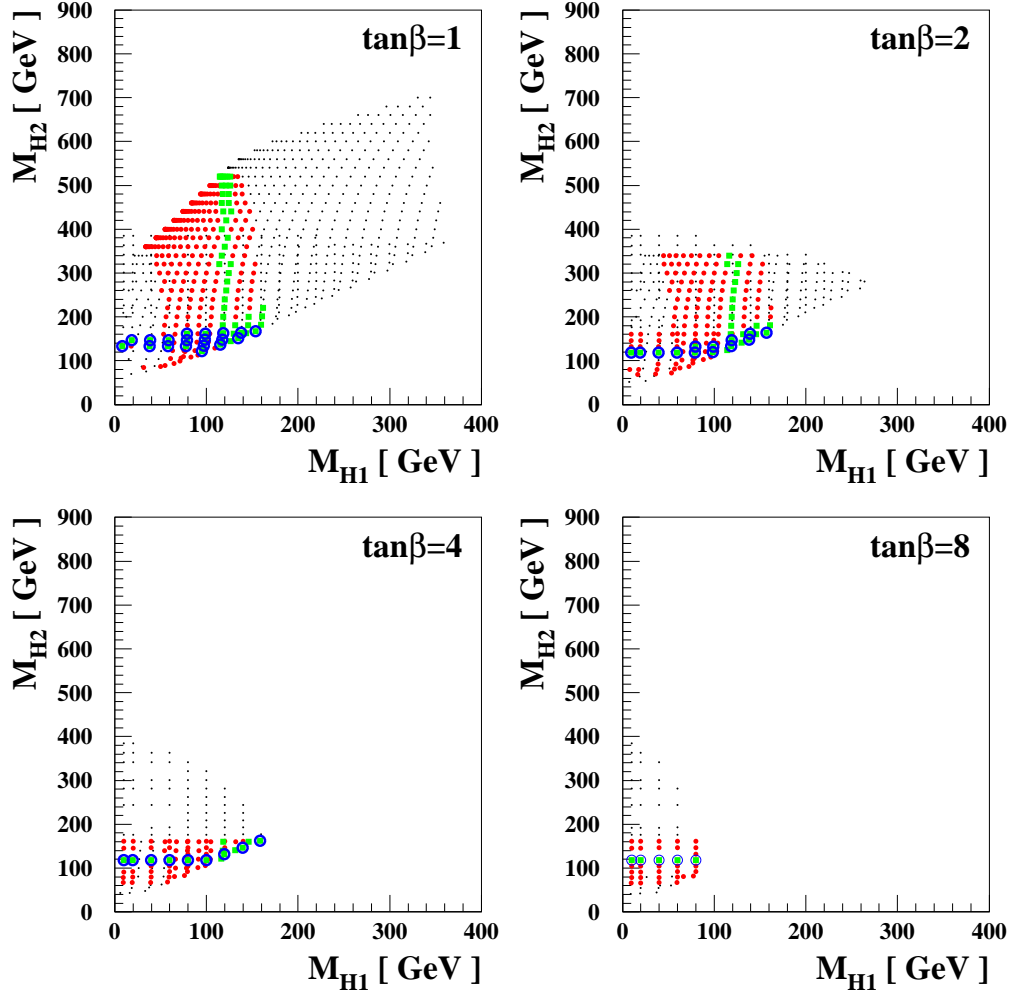,height=15.0cm,width=15.0cm}}
\end{center}
\vspace{-1.0cm}
\caption{\it The same as in Fig.~\ref{fig:mh12},
but with the restriction $M_{H^\pm}=M_{A}$.}
\label{fig:mh12p}
\end{figure}
\begin{figure}[!t]
\begin{center}
{\epsfig{figure=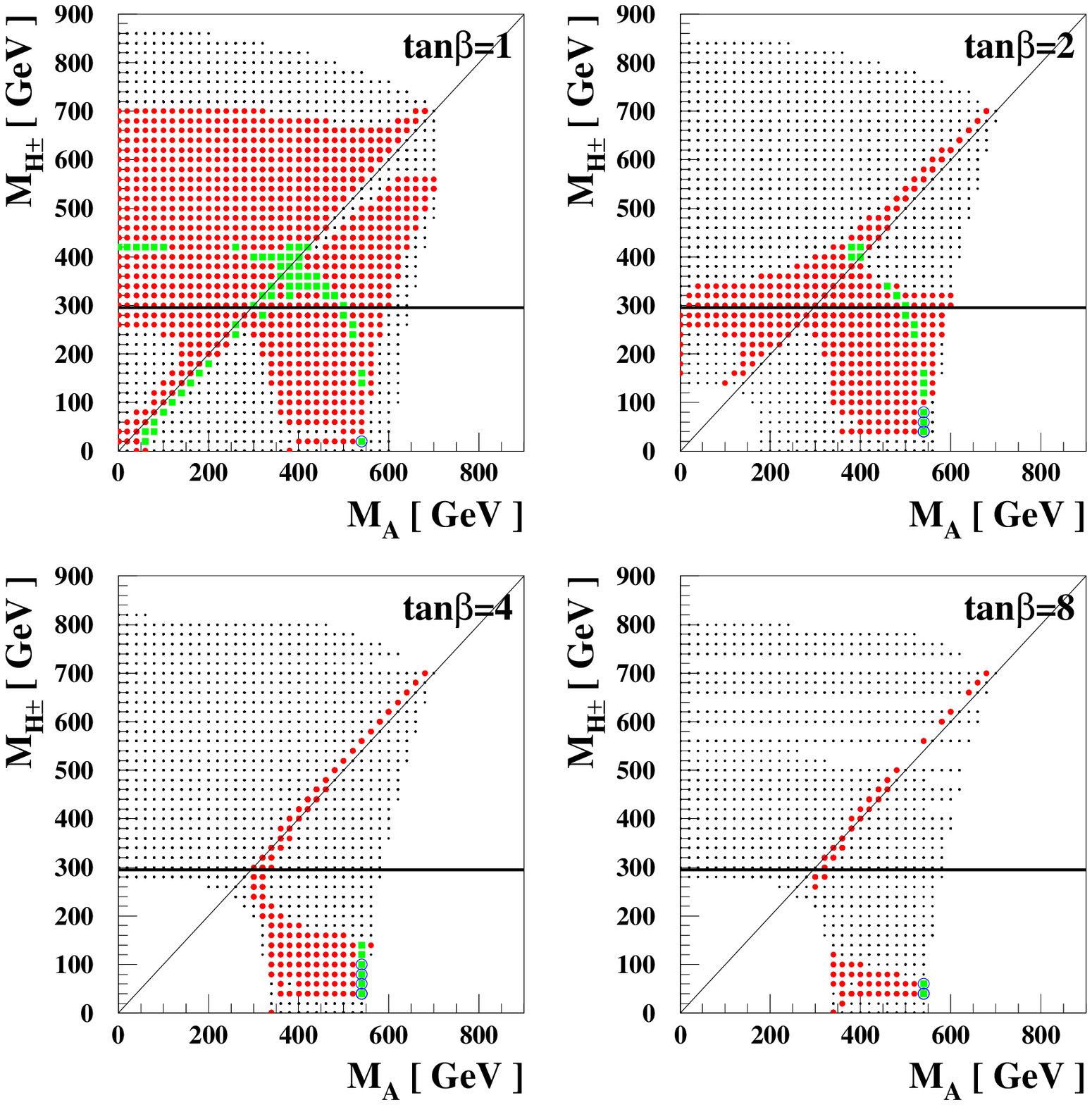,height=15.0cm,width=15.0cm}}
\end{center} 
\vspace{-1.0cm}
\caption{\it  The same  as in  Fig.~\ref{fig:mch_ma},  but restricting
 either $M_{H_1}$ or $M_{H_2}$ to lie between 123 and 127~GeV.}
\label{fig:mch_map}
\end{figure}

We  start our numerical  analysis by  showing in~Fig.~\ref{fig:mch_ma}
the  allowed parameter space  in the  $M_A-M_{H^\pm}$ plane,  which is
compatible  with   perturbative  unitarity  (black)   and  electroweak
precision limits (red) at the 90\% CL, for four values of $\tan\beta$:
$\tan\beta  = 1$  (upper left  panel),  $\tan\beta =  2$ (upper  right
panel), $\tan\beta = 4$ (lower  left panel) and $\tan\beta = 8$ (lower
right  panel).  Moreover,  the  green region  in~Fig.~\ref{fig:mch_ma}
indicates the  allowed area due  to the LEP  and LHC mass limits  on a
SM-like  Higgs boson.   The  little  blue circles  on  the green  area
highlight the region, governed  by the coupling hierarchy $|g_{H_2VV}|
> |g_{H_1VV}|$.  The thick horizontal  line that appears in each panel
of~Fig.~\ref{fig:mch_ma}  displays  the  lower  bound on  the  charged
Higgs-boson  mass $M_{H^\pm}\protect\gsim 295$  GeV, which  is derived
from the $b\to  s\gamma$ data~\cite{Misiak:2006zs}, assuming a Type-II
Yukawa coupling model.

From Fig.~\ref{fig:mch_ma},  we observe that  the combined constraints
get weaker  for low  values of $\tan\beta$,  with~$\tan\beta=1$ giving
the  weakest  exclusion  limits.    The  allowed  parameter  space  is
dominated  by   the  points  for  which   $M_{H^\pm}\approx  M_A$  and
$M_{H^\pm} \approx  M_{H_2}$ and centered  around~$400$~GeV.  This may
be understood as follows.  The direct constraints from LEP and the LHC
data restrict the mass of the SM-like Higgs boson to lie in the region
between 114.4~GeV  and 127~GeV.  This  is close to the  value 117~GeV,
for which $S_{\rm  SM}$ and $T_{\rm SM}$ almost  vanish.  On the other
hand,  the contributions  from the  heavier  Higgs bosons  to the  $T$
parameter  are significant,  unless  their masses  stay  close to  the
custodial   symmetric   limit,    where   $M_{H^\pm}   \approx   M_A$.
Alternatively,  an accidental  suppression of  the  $T_\Phi$ parameter
takes places,  when $M_{H^\pm}  \approx M_{H_2}$.  If  in view  of the
electroweak precision constraints we take $M_{H^\pm}=M_A=M_{H_2}\equiv
M_X$, then  the relation $M_{H_1}^2 \simeq  M_{\rm GW}^2 =  8 {\cal B}
v^2$ [cf.~(\ref{eq:mgw})] leads typically to
\begin{equation}
M_X^4\ \sim\ \frac{1}{4}\: \left(
8\pi^2 v^2 M_{H_1}^2-6 M_W^4 - 3 M_Z^4 +12 m_t^4 
\right)\; .
\end{equation}
Thus, for $M_{H_1} \sim 120$  GeV, one obtains an approximate estimate
of $M_X \sim 400$~GeV.

Let us now look more closely how each constraint acts on the parameter
space.  The  requirement of perturbative  unitarity (p.u.)  constrains
the masses of the charged and CP-odd Higgs bosons as follows:
\begin{equation}
M_{H^\pm}^{\rm p.u.}\ \lsim\ 850~{\rm GeV} \;, \qquad
M_A^{\rm p.u.}\ \lsim\ 700~{\rm GeV} \; .
\end{equation}
Note that  these upper bounds  are almost independent  of $\tan\beta$.
Instead, the  perturbative unitarity limit  on~$M_H$ depends crucially
on   $\tan\beta$,  which   becomes  stronger   for  large   values  of
$\tan\beta$.  This is a  direct consequence of the relation $\lambda_1
\simeq  M_H^2  t_\beta^2/2v^2$  and  the  perturbative  bound  imposed
on~$\lambda_1$.  Therefore, the  regions with small $M_{H^\pm}$ and/or
$M_A$  are excluded, since  $M_{H_1}^2$ gets  negative. The  reason is
that for $|g_{H_1VV}| >  |g_{H_2VV}|$, one has the relation $M_{H_1}^2
\simeq  M_{\rm GW}^2  = 8  {\cal B}  v^2$ and  the  one-loop parameter
${\cal B}$ given in~(\ref{eq:calB}) should be positive.

The electroweak (e.w.) oblique parameters offer additional constraints
on the scalar masses and on $\tan\beta$. Specifically, the mass limits
become stronger for larger values of $\tan\beta$, i.e.
\begin{eqnarray}
\tan\beta=1~: &&\quad 
M_{H^\pm}^{\rm p.u.\oplus e.w.} \lsim 700~{\rm GeV} \;, \qquad
M_A^{\rm p.u.\oplus e.w.} \lsim 700~{\rm GeV} \;,
\nonumber \\[2mm]
\tan\beta=2~: &&\quad
M_{H^\pm}^{\rm p.u.\oplus e.w.} \lsim 700~{\rm GeV} \;, \qquad 
M_A^{\rm p.u.\oplus e.w.} \lsim 700~{\rm GeV} \;,
\nonumber \\[2mm]
\tan\beta=4~: &&\quad
M_{H^\pm}^{\rm p.u.\oplus e.w.} \lsim 700~{\rm GeV} \;, \qquad 
300~{\rm GeV} \lsim M_A^{\rm p.u.\oplus e.w.} \lsim 700~{\rm GeV} \;,
\nonumber \\[2mm]
\tan\beta=8~: &&\quad
M_{H^\pm}^{\rm p.u.\oplus e.w.} \lsim 700~{\rm GeV} \;, \qquad 
300~{\rm GeV} \lsim M_A^{\rm p.u.\oplus e.w.} \lsim 700~{\rm GeV} \;,\qquad
\end{eqnarray}
where   the  superscript   ${\rm  p.u.\oplus   e.w.}$   indicates  the
simultaneous  implementation of limits  due to  perturbative unitarity
and the electroweak precision $S$ and $T$ parameters.

As a final constraint, we consider  the direct LEP and LHC limits on a
SM-like Higgs  boson mass. If we  combine these limits  with the bound
derived  on the  charged Higgs  mass $M_{H^\pm}\protect\gsim  295$ GeV
from the $b\to s\gamma$ data~\cite{Misiak:2006zs}, we find that
\begin{eqnarray}
\tan\beta=1~: &&\quad
295~{\rm GeV} \lsim M_{H^\pm} \lsim 680~{\rm GeV} \,, \qquad
M_A \lsim 650~{\rm GeV} \;, 
\nonumber \\[2mm]
\tan\beta=2~: &&\quad
295~{\rm GeV} \lsim M_{H^\pm} \lsim 600~{\rm GeV} \;, \qquad 
320~{\rm GeV} \lsim M_A \lsim 580~{\rm GeV} \; ,
\nonumber \\[2mm]
\tan\beta=4~: &&\quad
M_{H^\pm} \simeq M_A \sim 380 - 480~{\rm GeV}\; ,
\nonumber \\[2mm]
\tan\beta=8~: &&\quad
M_{H^\pm} \simeq M_A \sim 400 ~{\rm GeV}\; .
\end{eqnarray}
Finally,  it   is  worth  remarking  that  only   the  scenarios  with
$|g_{H_2VV}| > |g_{H_1VV}|$, which  are highlighted by blue circles in
the  plots, are  allowed  for larger  values  of $\tan\beta$,  e.g.~up
to~$\tan\beta =8$.

In Figure~\ref{fig:mh12},  we present  the allowed parameter  space in
the $M_{H_1}$--$M_{H_2}$ plane.  The allowed parameter space decreases
when $\tan\beta$ deviates from $1$.  When $\tan\beta=1$, we find there
exist three favourable mass regions:
\begin{eqnarray}
&\mbox{I.}& M_{H_1} > 127~{\rm GeV}~\hspace{1.45cm}:
M_{H_1}^{t_\beta=1} \sim 127 - 350~{\rm GeV} \;,  \ \
M_{H_2}^{t_\beta=1} \sim 140-380~{\rm GeV}\;,
\nonumber \\[2mm]
&\mbox{II.}&
M_{H_1} = 114 - 127~{\rm GeV} \ \ ~:
M_{H_1}^{t_\beta=1} = 114 - 127~{\rm GeV} \;, \ \
M_{H_2}^{t_\beta=1} \sim 140-550~{\rm GeV}\;,
\nonumber \\[2mm]
&\mbox{III.}& M_{H_1} <  114~{\rm GeV}~\hspace{1.45cm}:
M_{H_1}^{t_\beta=1} < 114~{\rm GeV} \;, \hspace{1.45cm}
M_{H_2}^{t_\beta=1} \sim 120-170~{\rm GeV}\;. \nonumber \\
\end{eqnarray}
In the  region I, the  mixing between the  $H_1$ and $H_2$  scalars is
significant  with  $M_{H_1} \sim  M_{H_2}$.   In  this  case, the  LHC
exclusion limits on a SM-like Higgs boson may not be straightforwardly
applicable. For  this reason, our  obtained limits should  be regarded
conservative in this case.  On the other hand, there is no lower limit
on the $H_1$  boson lying in the region~III  with $g_{H_1VV}^2 \ll 1$,
thus allowing for a very light scalar to have escaped detection at the
LEP~II collider.  For the larger values of $\tan\beta$, scenarios with
$g_{H_1VV}^2 \ll g_{H_2VV}^2$ are becoming more likely.  For~instance,
when $\tan\beta=8$, we find
\begin{equation}
M_{H_1}^{t_\beta=8}\ \lsim\ 80~{\rm GeV}\; , \qquad
M_{H_2}^{t_\beta=8}\ \sim\ 118~{\rm GeV}\;. \quad 
\end{equation}

Figure~\ref{fig:mh12p}  shows  the  allowed  parameter  space  in  the
$M_{H_1}$--$M_{H_2}$  plane,  for  the  custodial  symmetric  scenario
with~$M_{H^\pm}=M_{A}$.   As  explained  in the  previous  subsection,
$T_\Phi$    vanishes   identically    in   this    scenario,   because
$F_\Delta(M_A,M_{H^\pm})=0$    and    $\delta_Z^{H_i}=\delta_W^{H_i}$.
Therefore, the masses $M_{H_1}$ or $M_{H_2}$ must be close to 120 GeV,
in order  for the  SM contribution $T_{\rm  SM}$ to  remain acceptably
small.

Motivated  by the  2.3$\sigma$ excess  of a  positive SM  Higgs signal
corresponding to $M_{H_{\rm SM}} \sim 125$~GeV~\cite{LHC_SM_Higgs}, we
show   in  Figure~\ref{fig:mch_map}   the  allowed   regions   in  the
$M_A$--$M_{H^\pm}$  plane,  where   either  the  $H_1$-boson  or  the
$H_2$-boson   mass  is   restricted   to  lie   in  the   interval
$(123,127)$~GeV.  Taking  into account the lower bound  on the charged
Higgs-boson  mass, $M_{H^\pm}  \sim 295$~GeV,  derived from  $b  \to s
\gamma$ data, we find that all viable scenarios must have $|g_{H_1VV}|
> |g_{H_2VV}|$ and  $\tan\beta \leq 2$.   In this case, 
we find the following three possible scenarios:
\begin{itemize}
\item $M_{H_2}\sim M_A \sim 400$~GeV with $M_{H^\pm}  \lsim 420$~GeV
\item $M_A \lsim 100$~GeV with $M_{H^\pm}\sim  M_{H_2} \sim  400$ GeV
\item $M_{H_2} \lsim 180$~GeV with $M_{H^\pm}\sim  M_{A} \sim  400$ GeV
\end{itemize}
\noindent
In~conclusion,  if $M_{H_1} \sim
125$~GeV, viable  scenarios of the  SI-2HDM generically have  at least
two heavy Higgs bosons of $\sim 400$-GeV mass and favour low values of
$t_\beta\sim 1$.

\section{Conclusions}
\label{sec:conclusions}

We  have  studied the  Higgs  sector  of  a classical  scale-invariant
realization of  the two-Higgs-doublet  model (SI-2HDM).  Such  a model
may provide a minimal and  calculable solution to the well-known gauge
hierarchy  problem.    To  naturally  suppress   flavour  off-diagonal
interactions of the Higgs bosons  to quarks, we have imposed the usual
${\rm  Z}_2$ symmetry  on the  SI-2HDM  potential. In  this case,  the
SI-2HDM scalar  potential only depends  on the five  quartic couplings
$\lambda_{1-5}$ and hence it becomes very predictive.

The classical  scale symmetry of  the SI-2HDM is explicitly  broken by
quantum loop  effects due to gauge  interactions, Higgs self-couplings
and top-quark  Yukawa couplings. To~take account of  these effects, we
have  calculated the  one-loop effective  potential and  evaluated the
radiatively  corrected masses of  the CP-even  Higgs bosons  and their
mixing.   Unlike the  CP-even Higgs  sector,  we have  found that  the
CP-odd and  charged Higgs mass matrices retain  their tree-level form.
In  addition  to  the  CP-even  Higgs masses,  radiative  effects  may
drastically modify  the Higgs couplings  to the $Z$ boson,  through an
effective  $H_1$-$H_2$  mixing.   Our  analysis has  revealed  that  a
critical value  of the coupling $\lambda_3^{\rm c}$  exists, for which
$|g_{H_1VV}| = |g_{H_2VV}|$. Depending  on the value of $\lambda_3$, a
level-crossing phenomenon  occurs for both the $H_1$  and $H_2$ masses
and their couplings to the  $Z$ boson. For $\lambda_3 < \lambda_3^{\rm
  c}$, the lighter  state $H_1$ behaves like the  SM Higgs boson, with
$g_{H_1VV}^2  \sim  1$  and  its  mass is  well  approximated  by  the
Gildener-Weinberg mass $M_{H_1} \sim  M_{\rm GW}$, while $M_{H_2} \sim
M_H$.  Instead, if $\lambda_3  > \lambda_3^{\rm c}$, the heavier state
$H_2$  becomes SM-like  with  $g_{H_2VV}^2  \sim 1$  and  its mass  is
approximately given  by $M_{H_2}\sim M_{\rm GW}$,  while $M_{H_1} \sim
M_H$.

In our numerical analysis, we have imposed three basic theoretical and
phenomenological constraints on the  SI-2HDM: (i) the requirement of
validity of perturbative unitarity, (ii) the indirect constraints from
the  electroweak  precision data  and  (iii)  the direct  Higgs-search
constraints from the LEP collider  and the LHC.  At large $\tan\beta$,
the perturbative unitarity bounds  and the indirect constraints become
rather strong.   In conjunction with  the existing LEP and  the current
LHC limits  on the SM Higgs-boson mass,  the electroweak $T$-parameter
constraints reduce the theoretically  allowed parameter space into two
smaller regions, governed  by the approximate restrictions: $M_{H^\pm}
\sim M_A$ or  $M_{H^\pm} \sim M_{H_2}$. In this  context, our analysis
has shown that the Higgs-boson masses obey the following upper limits:
\begin{eqnarray}
M_{H_1} \lsim 350~{\rm GeV}\,, \ \
M_{H_2} \lsim 550~{\rm GeV}\,, \ \
M_{A} \lsim 650~{\rm GeV}\,, \ \
M_{H^\pm} \lsim 680~{\rm GeV}\,.
\nonumber
\end{eqnarray}
The  above bounds  hold for  low values  of $\tan\beta  \sim  1$.  For
$\tan\beta  \gsim  4$, the  masses  may  be  further restricted,  with
$M_{H^\pm}  \simeq M_A  \sim 400-500$  GeV. In  addition,  the heavier
CP-even state $H_2$  becomes more SM like with  $M_{H_2} \sim 114-170$
GeV and $M_{H_1}\lsim 160$ GeV.

Motivated by the 2.3$\sigma$ excess for a Higgs mass around 125~GeV at
the LHC, we  have extended our analysis by including  the bound on the
charged  Higgs mass  $M_{H^\pm}\protect\gsim 295$  GeV from  the $b\to
s\gamma$ data. In this case, we have found that $\tan\beta \sim 1$ and
the  lightest   Higgs  boson  is  SM   like,  with  $M_{H_1}=M_{H_{\rm
SM}}\simeq 125$  GeV.  The  heavier CP-even Higgs  boson $H_2$  can be
lighter than 180  GeV when $M_{H^\pm} \sim M_A \sim  400$ GeV.  On the
other hand,  the CP-odd scalar  $A$ can be  lighter than 100  GeV when
$M_{H^\pm}\sim M_{H_2} \sim 400$  GeV.  Otherwise, the pronounced mass
region  for $H_2$ and  $A$ is  mainly around  400 GeV  with $M_{H^\pm}
\lsim 420$  GeV.  We may  therefore conclude that, if  $M_{H_{\rm SM}}
\sim 125$~GeV, there  are at least two heavy  Higgs bosons with masses
close  to 400~GeV  and  the third  one  below $\sim  500$  GeV in  the
SI-2HDM.

At the LHC, the heavy neutral  Higgs bosons $H_2$ and $A$, with masses
$M_{H_2,A}\sim  400$~GeV,  are  expected  to be  mainly  produced  via
gluon--gluon fusion, where the  Higgs-pair production channel might be
also relevant. In general, the  search strategies for the Higgs bosons
$H_2$, $A$  and $H^\pm$ will depend  on the type of  the Yukawa sector
assumed.  Moreover, the detection  of possible light Higgs bosons with
masses  below $100$  GeV  and suppressed  couplings  to vector  bosons
becomes a  difficult issue.  A detailed investigation  of the possible
search strategies may be given elsewhere.

Another  problem that  needs  to  be addressed  in  detail within  the
SI-2HDM  pertains   the  natural  implementation   of  light  neutrino
masses. If  the theory is  extended with right-handed  neutrinos, then
light neutrino masses  can only be incorporated in the  theory in a SI
manner,  via  the standard  but  very  small  Dirac Yukawa  couplings.
However,  in  the presence  of  extra  singlets  or triplets,  further
possibilities  arise  to  naturally   explain  the  smallness  of  the
light-neutrino      masses,     along     the      lines     presented
in~\cite{Foot:2007ay,Meissner:2008gj,AlexanderNunneley:2010nw}.      It
would be interesting  to investigate the phenomenological implications
of such extensions of the SI-2HDM in a future communication.

\subsection*{Acknowledgements}

The  work  of  JSL  is  supported   in  part  by  the  NSC  of  Taiwan
(100-2112-M-007-023-MY3)    and    the    work    of   AP    by    the
Lancaster--Manchester--Sheffield  Consortium  for Fundamental  Physics
under the STFC grant ST/J000418/1.

\newpage

\def\theequation{\Alph{section}.\arabic{equation}}
\begin{appendix}

\setcounter{equation}{0}
\section{Vertex Corrections and Trilinear Higgs Couplings}\label{sec:vtx}

In this appendix we  calculate the one-loop quantum corrections ${\cal
  O}(\lambda^2_{1-5})$ to the  gauge-invariant, transverse part of the
gauge couplings  to neutral and  charged Higgs bosons.   These quantum
effects  get enhanced  for  large potential  couplings  and should  be
included  next to  the tree-level  contributions.  Our  calculation is
performed  in the  effective potential  limit, in  which  all external
momenta squared are assumed to vanish.

The   radiative    corrections   to   the    $Z$-$H^\pm$-$H^\mp$   and
$\gamma$-$H^\pm$-$H^\mp$ couplings are the  same. In~detail, these are
given by
\begin{eqnarray}
\delta_{Z}^{H^\pm} &=&
\delta_{\gamma}^{H^\pm} \equiv
\delta_{\gamma Z}^{H^\pm} =
\frac{v^2}{16\pi^2}
\sum_{j=1,2}
\lambda_{H_jH^-H^+}^2\,f_V(M_{H^\pm}^2,M_{H_j}^2,M_{H^\pm}^2)\; .
\end{eqnarray}
Here, $f_V(m_1^2,m_2^2,m_3^2)$ is  the one-loop vertex function, which
has been calculated to be
\begin{eqnarray}
f_V(m_1^2,m_2^2,m_3^2)
& =&  \frac{1}{(m_3^2-m_1^2)} 
\left[
\frac{m_3^2}{2(m_2^2-m_3^2)} -\frac{m_1^2}{2(m_2^2-m_1^2)}
\right. \nonumber \\[2mm]
&& \left. 
+\frac{m_3^4}{2(m_2^2-m_3^2)^2}\,\ln\left(\frac{m_3^2}{m_2^2}\right)
-\frac{m_1^4}{2(m_2^2-m_1^2)^2}\,\ln\left(\frac{m_1^2}{m_2^2}\right)
\right]\;, \nonumber
\end{eqnarray}
with $f_V(m^2,m^2,m^2)=1/(6m^2)$.   Likewise, the one-loop corrections
to the $H_i$-$A$-$Z$ couplings are given by
\begin{eqnarray}
\delta_Z^{H_i} &=& \frac{v^2}{16\pi^2}\left[
-\lambda_{H_iAA}\,
\sum_{j=1,2} g_{H_jAZ}\, \lambda_{H_jAA}\, f_V(M_A^2,M_A^2,M_{H_j}^2) \right.
\nonumber \\[2mm]
&& \left.
+ \sum_{(j,k)=(1,1)}^{(1,2),(2,1),(2,2)}
\lambda_{H_iH_jH_k}\, g_{H_jAZ}\, \lambda_{H_kAA} \,f_V(M_{H_j}^2,M_{H_k}^2,M_{A}^2)
\right] \; .
\end{eqnarray}
By analogy,  the one-loop  corrections to the  $A$-$H^\pm$-$W^\mp$ and
$H_i$-$H^\pm$-$W^\mp$ couplings are given by
\begin{eqnarray}
\delta_W^{A} & = & \frac{v^2}{16\pi^2}\left[
\sum_{j=1,2} \lambda_{H_jAA}\,\lambda_{H_jH^-H^+}
\,f_V(M_{A}^2,M_{H_j}^2,M_{H^\pm}^2)
\right]\,, \nonumber \\
\delta_W^{H_i} & = & \frac{v^2}{16\pi^2}\left[
-\lambda_{H_iH^-H^+}\,\sum_{j=1,2}
g_{H_jH^- W^+}\, \lambda_{H_jH^-H^+}\,f_V(M_{H^\pm}^2,M_{H^\pm}^2,M_{H_j}^2)
\right.
\nonumber \\[2mm]
&& \left.
+ \sum_{(j,k)=(1,1)}^{(1,2),(2,1),(2,2)}
\lambda_{H_iH_jH_k}\, g_{H_jH^- W^+}\, \lambda_{H_kH^-H^+}
\,f_V(M_{H_j}^2,M_{H_k}^2,M_{H^\pm}^2)
\right]\,.
\end{eqnarray}
Notice that $\delta_Z^{H_i}=\delta_W^{H_i}$ in the custodial symmetric
limit:  $M_A=M_{H^\pm}$ or  $\lambda_4=\lambda_5$,  since $\lambda_{H_i
AA}=\lambda_{H_i H^+H^-}$.

The Higgs potential terms  describing the trilinear Higgs interactions
may be written down as follows:
\begin{eqnarray}
V_{\rm Trilinear} &=& v\,\Bigg(
 \frac{\lambda_{_{H_1H_1H_1}}}{6}\,H_1^3\,
+\frac{\lambda_{_{H_1H_1H_2}}}{2}\,H_1^2H_2\,
+\frac{\lambda_{_{H_1H_2H_2}}}{2}\,H_1H_2^2\,
+\frac{\lambda_{_{H_2H_2H_2}}}{6}\,H_2^3\,
\nonumber \\
&&\hspace{0.3cm}
+\frac{\lambda_{_{H_1AA}}}{2}\,H_1 A A\,
+\frac{\lambda_{_{H_2AA}}}{2}\,H_2 A A\,
+\lambda_{_{H_1G^0A}}\,H_1 G^0 A\,
+\lambda_{_{H_2G^0A}}\,H_2 G^0 A\,
\nonumber \\
&&\hspace{0.3cm}
+\frac{\lambda_{_{H_1G^0G^0}}}{2}\,H_1 G^0 G^0\,
+\frac{\lambda_{_{H_2G^0G^0}}}{2}\,H_2 G^0 G^0\,\Bigg)
\nonumber \\ &+&
 v\,\sum_{i=1,2}\,\bigg[
 \lambda_{_{H_iG^-G^+}}\,H_iG^-G^+
+ \lambda_{_{H_iG^\mp H^\pm}}\,H_i(G^-H^++G^+H^-)
\nonumber \\ &&\hspace{1.1cm}
+ \lambda_{_{H_iH^- H^+}}\,H_iH^-H^+ \bigg]\; ,
\end{eqnarray}
where the trilinear self-couplings of the CP-even Higgs bosons are
\begin{eqnarray}
\lambda_{_{H_1H_1H_1}} &=& 6 \left(
 O_{11}^3\,\lambda_{_{\phi_1\phi_1\phi_1}}
+O_{11}^2O_{21}\,\lambda_{_{\phi_1\phi_1\phi_2}}
+O_{11}O_{21}^2\,\lambda_{_{\phi_1\phi_2\phi_2}}
+O_{21}^3\,\lambda_{_{\phi_2\phi_2\phi_2}} \right)\; ,
\nonumber \\[3mm]
\lambda_{_{H_1H_1H_2}} &=&
6\,O_{11}^2O_{12}\,\lambda_{_{\phi_1\phi_1\phi_1}}
+2\,(O_{11}^2O_{22}+2\,O_{11}O_{12}O_{21})\,\lambda_{_{\phi_1\phi_1\phi_2}}
\nonumber \\  && \hspace{2.9cm}
+2\,(O_{12}O_{21}^2+2\,O_{11}O_{21}O_{22})\,\lambda_{_{\phi_1\phi_2\phi_2}}
+6\,O_{21}^2O_{22}\,\lambda_{_{\phi_2\phi_2\phi_2}}\; ,
\nonumber \\[3mm]
\lambda_{_{H_1H_2H_2}} &=&
6\,O_{11}O_{12}^2\,\lambda_{_{\phi_1\phi_1\phi_1}}
+2\,(O_{12}^2O_{21}+2\,O_{11}O_{12}O_{22})\,\lambda_{_{\phi_1\phi_1\phi_2}}
\nonumber \\  && \hspace{2.9cm}
+2\,(O_{11}O_{22}^2+2\,O_{12}O_{21}O_{22})\,\lambda_{_{\phi_1\phi_2\phi_2}}
+6\,O_{21}O_{22}^2\,\lambda_{_{\phi_2\phi_2\phi_2}}\; ,
\nonumber \\[3mm]
\lambda_{_{H_2H_2H_2}} &=& 6 \left(
 O_{12}^3\,\lambda_{_{\phi_1\phi_1\phi_1}}
+O_{12}^2O_{22}\,\lambda_{_{\phi_1\phi_1\phi_2}}
+O_{12}O_{22}^2\,\lambda_{_{\phi_1\phi_2\phi_2}}
+O_{22}^3\,\lambda_{_{\phi_2\phi_2\phi_2}} \right)\; .
\end{eqnarray}
In addition, the trilinear couplings involving one CP-even Higgs boson
and two CP-odd scalars may be cast into the form:
\begin{eqnarray}
\lambda_{_{H_iXY}}\ =\ N_{XY}\, (
 O_{1i}\lambda_{_{\phi_1XY}}
+O_{2i}\lambda_{_{\phi_2XY}}) \;, \qquad
\end{eqnarray}
with $(XY\,,N_{XY})=(AA\,,2)\,,\ (G^0A\,,1)\,,\ (G^0G^0\,,2)$.

Finally, the trilinear CP-even  Higgs couplings with the charged Higgs
bosons $H^\pm$ may be expressed as follows:
\begin{eqnarray}
\lambda_{_{H_iX^\prime Y^\prime }}\ =\
 O_{1i}\lambda_{_{\phi_1X^\prime Y^\prime }}
+O_{2i}\lambda_{_{\phi_2X^\prime Y^\prime }}\; ,
\end{eqnarray}
with  $X^\prime  Y^\prime=G^-G^+$, $G^\mp  H^\pm$  and $H^-H^+$.   The
trilinear couplings in the basis of weak eigenstates are given by
\begin{eqnarray}
&&
\lambda_{_{\phi_1\phi_1\phi_1}} = \lambda_1 c_\beta\,, \ \ \
\lambda_{_{\phi_1\phi_1\phi_2}} = \frac{\lambda_{345}}{2} s_\beta\,, \ \ \
\lambda_{_{\phi_1\phi_2\phi_2}} = \frac{\lambda_{345}}{2} c_\beta\,, \ \ \
\lambda_{_{\phi_2\phi_2\phi_2}} = \lambda_2 s_\beta\,;
\nonumber \\[0.3cm]
&&
\lambda_{_{\phi_1AA}} = \lambda_1c_\beta s_\beta^2 +\frac{\lambda_{34}}{2} c_\beta^3
-\frac{\lambda_5}{2}c_\beta (1+s_\beta^2) \,, \ \ \
\lambda_{_{\phi_2AA}} = \lambda_2s_\beta c_\beta^2 +\frac{\lambda_{34}}{2} s_\beta^3
-\frac{\lambda_5}{2}s_\beta (1+c_\beta^2)\,;
\nonumber \\[0.3cm]
&&
\lambda_{_{\phi_1G^0A}} = (-2\lambda_1+\lambda_{34})c_\beta^2s_\beta -
\lambda_5 s_\beta^3 \,, \ \ \ 
\lambda_{_{\phi_2G^0A}} = (2\lambda_2-\lambda_{34})s_\beta^2c_\beta + \lambda_5 c_\beta^3
\,; 
\nonumber \\[0.3cm] &&
\lambda_{_{\phi_1G^0G^0}} = \lambda_1 c_\beta^3+\frac{\lambda_{345}}{2}c_\beta s_\beta^2
\,, \ \ \
\lambda_{_{\phi_2G^0G^0}} = \lambda_2 s_\beta^3+\frac{\lambda_{345}}{2}s_\beta c_\beta^2
\\[1.0cm] &&
\lambda_{_{\phi_1G^-G^+}}=2\lambda_1 c_\beta^3+\lambda_{345}s_\beta c_\beta^2\,, \ \ \
\lambda_{_{\phi_2G^-G^+}}=2\lambda_2 s_\beta^3+\lambda_{345}c_\beta s_\beta^2\,,
\nonumber \\[0.3cm] &&
\lambda_{_{\phi_1G^\mp G^\pm}}=-2\lambda_1 s_\beta c_\beta^2+\lambda_3 s_\beta c_\beta^2
+\frac{\lambda_{45}}{2}\,s_\beta c_{2\beta}\,, \ \ \
\lambda_{_{\phi_2G^\mp G^\pm}}=2\lambda_2 c_\beta s_\beta^2-\lambda_3 c_\beta s_\beta^2
+\frac{\lambda_{45}}{2}\,c_\beta c_{2\beta}\,, \ \ \
\nonumber \\[0.3cm] &&
\lambda_{_{\phi_1H^-H^+}}=2\lambda_1 c_\beta s_\beta^2+\lambda_3 c_\beta^3
-\lambda_{45}c_\beta s_\beta^2\,, \ \ \
\lambda_{_{\phi_2H^-H^+}}=2\lambda_2 s_\beta c_\beta^2+\lambda_3 s_\beta^3
-\lambda_{45}s_\beta c_\beta^2\,, \ \ \
\end{eqnarray}
with     $\lambda_{345}\equiv    \lambda_3+\lambda_4+\lambda_5$    and
$\lambda_{34}\equiv       \lambda_3+\lambda_4$.       Notice      that
$\lambda_{\phi_i  H^+H^-}  = 2\lambda_{\phi_i  AA}$  in the  custodial
symmetric limit: $\lambda_4=\lambda_5$.

\end{appendix}


\end{document}